# FLUID-STRUCTURE INSTABILITY FORECASTS THORACIC AORTIC ANEURYSM PROGRESSION

TOM Y. ZHAO[†,⊖,§] | PHD, ETHAN M.I. JOHNSON[‡,⊖] | PHD, GUY ELISHA[†], SOURAV HALDER[†], BEN C. SMITH [∝], BRADLEY D. ALLEN [⊕] | MD, MICHAEL MARKL[‡,⊕] | PHD, AND NEELESH A. PATANKAR[†,§] | PHD

The basic mechanism driving aneurysm growth is unknown. Currently, clinical diagnosis of an aneurysm is mainly informed by retrospective tracking of its size and growth rate. However, aneurysms can rupture before reactive criteria are met or remain stable when they are exceeded. Here, we identify a fluid-structure instability that is associated with abnormal aortic dilatation. Our analysis yields a measurable dimensionless number and its analytically derived critical threshold. This threshold pinpoints the transition from stable flow to unstable aortic fluttering as a function of the physiological properties composing the dimensionless number, like blood pressure and aortic compliance. A retrospective study was then conducted with 4D-flow MRI data from 117 patients indicated for cardiac imaging and 100 healthy volunteers recruited prospectively. The difference between the dimensionless number and its critical threshold was calculated for every subject from their earliest MRI data and used as an aneurysm physiomarker to forecast future growth. As a binary predictor for abnormal growth and subsequent surgical intervention reported from follow-up imaging, the aneurysm physiomarker yielded an AUC of 0.997 in a receiving operator characteristic analysis. Though validated here for thoracic ascending aortic aneurysms, this instability mechanism may be used to understand, predict and inform patient-specific treatment of aneurysms in any location without fundamental differences.

## 1. Introduction

Aneurysms are pathological, localized dilations of a blood vessel that may occur throughout the human body. Intracranial, thoracic aortic, and abdominal aortic aneurysms (IA, TAA, AAA) are each estimated to occur with a global prevalence of $2-5\%$ [1;2;3]. Rupture of an aneurysm induces a high rate of mortality and morbidity for the patient. Studies showed that over half of patients with ruptured TAAs or AAAs died before reaching a hospital, with overall mortality ranging from 80 to 100% [4;5]. For patients with IA, between 10 to 30% died suddenly away from hospitals [3], and of those admitted for treatment, 45% experienced an outcome categorized as either moderately disabled, severely disabled, vegetative survival, or death on the Glasgow Outcome Disability Scale. Surgical intervention can be performed to prevent rupture but also carries the risk of complications or death [4]. Thus, it is vital to accurately predict the risk of aneurysm formation and abnormal aortic growth to inform timely treatment.

### 1.1. **Current standard of care.**

The standard of care is to recommend elective treatment for aneurysms based on correlations between rupture risk and aneurysm dimensions. For TAAs, the chance of rupture increases from 2% for diameters between 4 and 4.9 cm to 7% for diameters above 6 cm [6]. The mean growth rate is approximately 0.1 cm/year [7]. This informs current clinical practice, which suggests surgical

(†) Northwestern University, Department of Mechanical Engineering: 2145 Sheridan Road, Evanston, Illinois 60208, USA

(‡) Northwestern University, Department of Biomedical Engineering: 2145 Sheridan Road, Evanston, Illinois 60208, USA

(⊕) Northwestern University, Department of Radiology: 676 N St Clair St, Chicago, IL 60611, USA

(∝) Northwestern University Feinberg School of Medicine, 420E Superior St, Chicago, IL 60611, USA

(⊖) These authors contributed equally to this work

(§) Corresponding authors: n-patankar@northwestern.edu ; tomzhao@u.northwestern.edu

intervention for aneurysm diameters larger than a range between 5.5 to 6.0 cm or exhibiting growth rate larger than a range between 0.5 to 1 cm/year, depending on the aneurysm location and patient history [7;8]. However, clinical assessment of growth requires comparison between images taken at two time points, typically between 2 to 5 years. Over this period, an aneurysm can grow significantly or rupture fatally. Conversely, an aneurysm which exceeds these statistical criteria may nonetheless remain stable. Thus, prevailing diagnostic guidelines are retrospective and apply population trends to individual patients. To improve predictive capability, the fundamental mechanism underlying aneurysm growth, dissection and rupture must be resolved.

1.2. **Tissue mechanics associated with aneurysm progression.** A literature review of clinical observations on how aneurysm distensibility evolves during disease progression is provided in Table S3. As an aneurysm enlarges, the aortic wall degrades due to the loss of elastin and smooth muscle content (SMC) [9;10;11]. This stage, sometimes also referred to as Stage 1, has been reported to be accompanied by intimal thickening [12]. Overall, the wall stiffness is found to decrease in this stage [12].

In the subsequent stage (Stage 2), further decrease in elastin and SMC content has been reported along with the formation of a neo-adventitia layer from new collagen deposition on the outer walls. Wall stiffness has been reported to further decrease [12].

Two possible developmental paths diverge after Stage 2. In Stage 3, either increasing collagen deposition stiffens the aortic wall to preclude further growth (a Type 1 aneurysm group), or the wall remodels to a weakened state due to a failure to lay down collagen, wall inflammation, and/or adipocyte accumulation (a Type 2 aneurysm group) [12]. This second branch triggers further growth and can lead to eventual dissection or rupture [11]. It is noted that among all the stages of disease progression, listed above, stiffening of the wall is reported in the Stage 3, Type 1 aneurysm group, whereas wall stiffness is reported to decrease in all other stages.

While this important body of work explicates the tissue mechanics underlying growth, the invasive biopsy required to characterize aortic wall makeup preclude its use in clinical decision-making.

1.3. **Prediction of aneurysm growth and rupture.** The pursuit of a causative relation between aneurysm progression and certain physical properties falling outside a normative range remains inconclusive. For instance, high blood pressure [13], abnormal wall shear stress distribution [14], large aortic size [6], and high wall compliance [12] have all been correlated with aneurysmal growth. However, it is uncertain how these factors interact to trigger abnormal aortic dilatation. For example, high shear stresses have been implicated in some scenarios while low shear stresses in other scenarios [14].

Thus, the state-of-the-art to predict aneurysm growth is based on regression analyses for risk factors such as age or smoking history [15]; regression on morphologic features such as aneurysm diameter or undulation index [16]; machine learning approaches trained on imaging features such as aneurysm diameter or intraluminal thrombi thickness [17]. These methods are based on establishing a correlation between available clinical data and aneurysm growth rates. As with all regression techniques, the breadth of data used to train the model is the main determinant for performance; with a small training cohort relative to the disease population, the predictive capability of the model becomes extrapolative rather than interpolative.

1.4. **A unifying hypothesis for aneurysm prediction.** Here, we introduce a unifying, ab initio hypothesis that elucidates the role of known physical factors – blood pressure, aortic size, wall shear stress, and pulse wave velocity – in the development of an aneurysm. The key ansatz is that when these interacting physiological variables fall outside of a normative range, they can trigger a fluid-structure instability that may lead to or signal the onset of abnormal aortic growth.

The dominant properties that destabilize the coupled fluid-structure motion within the aorta are the pressure gradient driving blood flow and the blood vessel diameter. They cause the vessel wall to 'flutter' under higher frequency, oscillatory modes of the heartbeat cycle. Concurrently, the viscosity dampens and the wall stiffness constrains these flutter perturbations to help stabilize the blood vessel. A first principles analysis of these competing factors yields a clinically measurable, dimensionless number that describes the transition from stable flow to unstable aortic fluttering. This is analogous to how the Reynolds number describes the transition from laminar to turbulent flow.

A physically intuitive analogy is the unstable fluttering of a banner in the wind, where the flow velocity, banner size, drag coefficient, and material elasticity take the place of blood pressure, aortic size, wall shear stress, and pulse wave velocity, respectively. Note that the pulse wave velocity, to be formally defined later in the paper, depends on material elasticity. Flutter in this mechanical context induces a significant increase in stresses within the material due to large deformations. Analogously, we hypothesize that the instability that induces aortic wall fluttering may lead to or signal the necessary conditions for aneurysm growth and eventual rupture.

Pulsatile flow in a compliant channel has been studied prior [18;19;20;21], in which the walls of a 2D channel are modeled as spring and damper backed plates. The main instabilities resolved are boundary shear flow instabilities such as the Tollmien-Schlichting wave, which drives the transition to turbulence. Elastic wall deformation is obtained via a Kelvin-Helmholtz type shear instability driven by the Stokes layer near the wall.

In this work, our focus is on what mechanisms act on the aortic wall to trigger aneurysm development and progression. We therefore resolve a tubular 1D fluid-structure instability that depends on flow pulsatility, wall shear, blood pressure, and pulse wave velocity (wall stiffness). The wall fluttering stemming from this instability is primarily pressure mediated via the tube law describing the behavior of the elastic tube. We find that this instability appears strongly correlated with abnormal aortic dilatation.

1.5. **Application of the instability-based aneurysm physiomarker.** This paper presents a theoretical analysis of the fluid-structure interaction that yields a critical threshold for the dimensionless number beyond which the instability occurs. This criticality condition is obtained from first principles and can be measured for each patient. Together, the dimensionless number minus the critical threshold encapsulates the instability onset potentially driving or signaling aneurysm progression. We further propose that this flutter instability parameter (dimensionless number minus critical threshold) can act as a aneurysm physiomarker to forecast abnormal aortic dilatation.

In a retrospective study of patients indicated for cardiac imaging with follow-up assessment of aortic dimensions available, we observed that the proposed aneurysm physiomarker is highly predictive of whether an aneurysm exhibits abnormal vs natural growth. The only input to calculate this aneurysm physiomarker for each patient is a single 4D flow magnetic resonance imaging (MRI) scan taken at an initial time point. This analytical determination was then compared with the clinical outcomes reported from a follow-up at least one year after the baseline MRI to evaluate its potential for predicting significant aortic dilation. As a binary predictor for abnormal growth and surgical intervention, the area under the curve (AUC) for a receiver operating characteristic analysis is 0.997. No training data is necessary to tune the calculation or performance of the aneurysm physiomarker.

The aneurysm physiomarker clarifies the exact interaction between physical properties like blood pressure and wall stiffness that trigger the instability and associated abnormal growth. Thus, it also reveals what physiological variables must be controlled to prevent this flutter instability. At a macro level, the dominant factor driving aneurysm progression is shown to vary depending on the subjects' aneurysm stage, which is useful for overall disease progression analysis. Patient-level differences are also captured explicitly by the aneurysm physiomarker, which can show the specific location along the aorta at highest risk for abnormal growth. Lastly, by binning subjects according to age and sex, we also found that the proposed aneurysm physiomarker dominantly describes the clinically observed population traits of aneurysm development in both patient and normal subject cohorts.

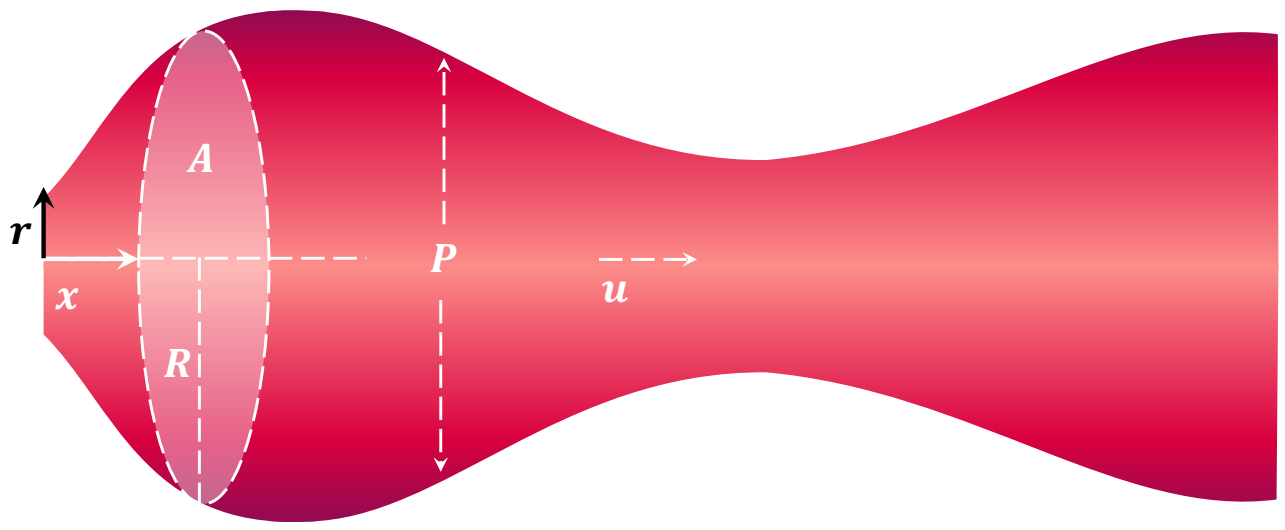

FIGURE 1. The distensible blood vessel is modeled as a one-dimensional system with excess internal pressure $P$ (normalized by density) and velocity $u$ being averaged across the radial direction $r$, which is normal to the centerline coordinate $x$. The interior area $A = \pi R^2$ varies as a function of both space $x$ and time $t$.

## 2. Theory

**2.1. Derivation of the ab initio aneurysm physiomarker.** Here, we derive the flutter instability parameter from first principles. A classical model for flow through a blood vessel consists of 1D conservation equations for mass and momentum from the Navier-Stokes equations, closed by a constitutive 'tube law' for the variation of pressure with the cross-sectional area [22;23] due to elasticity of the wall. The pressure gradient is chosen to vary periodically in time with frequency equal to that of the heartbeat cycle [24].

Following this problem formulation (Fig. 1), we conduct a linear stability analysis to determine a critical threshold beyond which the blood vessel area fluctuates unboundedly under infinitesimal perturbations. The blood vessel is assumed to be infinitely long along the axial direction to keep the theoretical analysis tractable. The base flow is chosen to be a periodic limit cycle following the pulsatile waveform of blood pressure over the cardiac cycle. The effect of perturbations at all higher order frequencies are resolved via the Floquet theory. We find that a single dimensionless number and its critical threshold describes the onset of the proposed instability which triggers the fluttering of the vessel wall.

2.1.1. *Governing equations.* In 1D, the mass and momentum conservation equations are [22;23]

$$\frac{\partial A}{\partial t} + \frac{\partial (uA)}{\partial x} = 0, \tag{1}$$

$$A\frac{\partial u}{\partial t} + \hat{\alpha} A u \frac{\partial u}{\partial x} = -A\frac{\partial P}{\partial x} + 2\pi \frac{R}{\rho}\tau_w, \tag{2}$$

where $A[x,t]$ and $R[x,t]$ denote the cross-sectional area and radius, while the pressure $P[x,t]$ and velocity $u[x,t]$ represent values averaged over the radial profiles at each location $x$ and time $t$. Here, $P$ is the excess internal pressure inside the blood vessel *normalized by the blood density* $\rho$. The wall shear stress term is $\tau_w$ and $\hat{\alpha}$ is a constant factor that arises from cross-sectional averaging of the non-linear convection term. Here, we take $\hat{\alpha} = 1$ [22;23].

To close the problem, the tube law relating pressure to area is taken to be linear [25]

$$P = \frac{K_e}{\rho}\left(\frac{A}{A_o} - 1\right), \tag{3}$$

where $K_e$ is the blood vessel wall stiffness and $A_o$ is the relaxed area of the blood vessel corresponding to excess internal pressure $P = 0$. In Supplementary S1.2, we show that the linear stability problem generalizes to any arbitrary tube law relating pressure to vessel area.

2.1.2. *Base flow.* For pulsatile blood flow, the base equilibrium solutions for area $A_b$, pressure gradient $\frac{\mathrm{d}P_b}{\mathrm{d}x}$, and velocity $u_b$ can be written as

$$A_b = A_m + A_\omega[t] \simeq A_m, \tag{4}$$

$$u_b = u_m + u_\omega[t] = u_m + \frac{1}{2}(\bar{u}_\omega e^{i\omega t} + \bar{u}_\omega^* e^{-i\omega t}), \tag{5}$$

$$-\frac{\partial P_b}{\partial x} = \phi_b = \phi_m + \phi_\omega[t] = \phi_b + \frac{\bar{\phi}_\omega}{2}(e^{i\omega t} + e^{-i\omega t}), \tag{6}$$

where $\omega$ is the angular frequency of the heartbeat cycle. $A_m$, $u_m$, and $\phi_m$ are the temporal mean values of area, velocity, and pressure gradient, respectively. $u_\omega$ and $\phi_\omega$ are the time dependent, oscillatory components. $\bar{u}_\omega$ is a complex amplitude associated with $u_\omega$ and superscript $*$ denotes the complex conjugate. The amplitude $\bar{\phi}_\omega$ associated with $\phi_\omega$ is taken to be real for simplicity since it is the driving term. Finally, note that for the given form of the driving pressure gradient $\frac{\partial P_b}{\partial x}$, the pressure $P_b$ and consequently the area $A_b$ (via the tube law) will vary along the axial ($x$) direction. Such variations in the base flow are typically on the order of 5% of the mean value as measured via transthoracic echocardiogram[26]. Thus in this work, we assume for simplicity that the area is approximately constant in the base state; that is, $A_b \simeq A_m$.

The base flow is then described by the conservation equations

$$\frac{\partial u_b}{\partial x} = 0, \tag{7}$$

$$A_m \frac{\partial u_\omega[t]}{\partial t} = A_m(\phi_m + \phi_\omega[t]) - \beta_m \pi \nu u_m - \beta_b \pi \nu u_\omega[t]. \tag{8}$$

For constant forcing (as opposed to a pulsatile flow), a parabolic velocity profile generates a corresponding wall shear stress $\tau_{w,\text{parabolic}} = \frac{-4\rho\nu u}{R}$, where the kinematic viscosity of blood is given by $\nu$ and the negative sign indicates that the wall shear stress on the fluid is pointed in the direction opposite to that of $u$. For the constant mean flow $u_m$ we assume parabolic flow based mean shear stress which is equivalent to $\beta_m = 8$ in eqn. 8. For the superposed oscillatory flow driven by the heartbeat cycle, the corresponding wall shear stress is obtained from a wall shear coefficient $\beta_b$ in the momentum equation via[22]

$$\beta_b[w_0] = 8\frac{\tau_w}{\tau_{w,\text{parabolic}}} = -2w_0 i^{3/2}\left(\frac{J_1[w_0 i^{3/2}]}{J_0[w_0 i^{3/2}]}\right)\left(\frac{1}{1 - 2\frac{J_1[w_0 i^{3/2}]}{J_0[w_0 i^{3/2}]}/(w_0 i^{3/2})}\right), \tag{9}$$

where $J_n$ denote Bessel functions of the first kind. The complex $\beta_b$ represents the ratio of wall shear stress (WSS) at a Womersley number $w_0 = R\sqrt{\omega/\nu} \geq 0$ (pulsatile flow driven at angular frequency $\omega$) to the fully developed WSS associated with $w_0 = 0$ (constant forcing). The factor $\beta_b$ is determined via the functional relationship between the wall shear stress $\tau_w$ on $w_0$ as derived by Womerlsey[27] (eqn. 9) and displayed in Fig. S1.

Finally, the mean terms $\phi_m$ and $u_m$ are related through momentum conservation (eqn. 8) via $u_m = \frac{\phi_m A_m}{\beta_m \pi \nu}$. Analogously, the oscillatory flow components are related by

$$\bar{u}_\omega = \frac{\bar{\phi}_\omega A_m(\beta_b \pi \nu - i\omega A_m)}{(\beta_b \pi \nu)^2 + (\omega A_m)^2}. \tag{10}$$

2.1.3. *Linearized perturbation equations.* Next, the base solutions for the velocity, area, and pressure are perturbed by infinitesimal quantities $Y'$ of the respective variables

$$Y = Y_b + Y' = Y_b + \sum_{k=-\infty}^{\infty} Y_k'[t] e^{ikx}, \tag{11}$$

for $Y \in \{A, u, P\}$. The perturbations $Y'$ are expressed as the sum of contributions from all wavenumbers $k$. After linearizing of the governing equations and subtracting the base solution, the equations for perturbation components $Y'_k[t]$ corresponding to wavenumber $k$ are

$$\frac{\partial A'_k}{\partial t} + A_b iku'_k + u_b ikA'_k = 0, \tag{12}$$

$$A_b \frac{\partial u'_k}{\partial t} + A'_k \frac{\partial u_b}{\partial t} + A_b u_b iku'_k = -A_b ikP'_k - \beta_b \pi \nu u'_k + \phi_b A'_k, \tag{13}$$

$$P'_k = \frac{K_e}{\rho} \frac{A'_k}{A_o}. \tag{14}$$

We can combine the tube law (eqn. 14) into the momentum equations (eqn. 13) to express pressure $P'_k$ in terms of area $A'_k$ perturbations. The complex valued solution set tightens to $\underline{X}_k = [A'_k, u'_k]^T$.

The perturbation equations (eqns. 12, 13) can be written in matrix form as

$$\dot{\underline{X}}_k = \boldsymbol{H}\underline{X}_k, \tag{15}$$

where $\dot{\underline{X}}_k$ denotes the time derivative. The vector $\underline{X}_k \in C(\mathbb{R}, M_{2,1}[\mathbb{C}])$, denoting a continuous, functional mapping from a real scalar in time to the complex vector space for $[A'_k, u'_k]^T$. The coefficient matrix $\boldsymbol{H}[t] \in C(\mathbb{R}, M_{2,2}[\mathbb{C}])$ is periodic with associated frequency $\omega$. This class of periodic linear systems under parametric forcing admits solutions of the Floquet form.

2.1.4. *Floquet solution.* The basis for all solutions to eqn. 15 can be expressed as the product of a periodic component, and an exponential term in time (Theorem 4.1 in Coddington et al.[28]). That is, $\boldsymbol{X}_k = \boldsymbol{P}(t)e^{\boldsymbol{R}t}$, where $\boldsymbol{P}[t]$ is invertible, $\boldsymbol{P}[t] = \boldsymbol{P}[t + 2\pi/\omega]$, and $\boldsymbol{P}[t] \in C(\mathbb{R}, M_{2,2}[\mathbb{C}])$. $\boldsymbol{X}_k[t] \in C(\mathbb{R}, M_{2,2}[\mathbb{C}])$ is a complex valued matrix formed from the fundamental solution to eqn. 15. To assess the stability of the solutions $\boldsymbol{X}_k$, we observe that the eigenvalues $\lambda$ of $\boldsymbol{R}$ determine the stability of the system. Specifically, if there exist $\lambda = \mu + i\alpha\omega$ such that $\mu < 0$ for all wavenumbers $k$, then the perturbations $A'$ and $u'$ decay in time. Otherwise if $\mu > 0$ for any wavenumber $k$, the base solution is unstable, which according to our hypothesis may trigger aneurysm formation and abnormal growth[28].

To find $\lambda$, we see that each solution vector of the fundamental solution takes the form $\underline{X}_k = e^{\lambda t}\underline{P}[t]$, where $\underline{P}[t]$ is a vector polynomial with coefficients periodic in the associated frequency $\omega$ (Section 4.5 in Coddington et al.[28]). This periodic function $\underline{P}[t]$ can therefore be written as the sum of temporal Fourier modes[29], such that $\underline{X}_k$ becomes

$$\underline{X}_k = \sum_{-\infty}^{\infty} \hat{\underline{X}}_{k,n} e^{(\mu + i(n+\alpha)\omega)t}, \tag{16}$$

$$\frac{\partial \underline{X}_k}{\partial t} = \sum_{-\infty}^{\infty} (\mu + i(n+\alpha)\omega)\hat{\underline{X}}_{k,n} e^{(\mu + i(n+\alpha)\omega)t}. \tag{17}$$

Using this in the linearized perturbation equations (eqns. 12, 13), we obtain

$$(\mu + i(n+\alpha)\omega)\hat{A}_{k,n} + A_b ik\hat{u}_{k,n} + u_b ik\hat{A}_{k,n} = 0, \tag{18}$$

$$A_b(\mu + i(n+\alpha)\omega)\hat{u}_{k,n} + \hat{A}_{k,n}\frac{\partial u_b}{\partial t} + A_b u_b ik\hat{u}_{k,n} = -A_b ik \frac{K_e}{\rho}\frac{\hat{A}_{k,n}}{A_o} - \beta_b \pi \nu \hat{u}_{k,n} + \phi_b \hat{A}_{k,n}. \tag{19}$$

Note that each equation corresponds to one temporal Fourier mode ($n\omega$, where $n \in \{-\infty, \infty\}$) at a particular spatial wavenumber $k$. We substitute the base solutions (eqns. 5, 6) to obtain the final homogeneous equation for the Fourier coefficients $\hat{\underline{X}}_{k,n} = [\hat{A}_{k,n}, \hat{u}_{k,n}]^T$.

$$(\mu + i(n+\alpha)\omega)\hat{A}_{k,n} + A_b ik\hat{u}_{k,n} + u_m ik\hat{A}_{k,n} + \frac{1}{2}\bar{u}_\omega ik\hat{A}_{k,n-1} + \frac{1}{2}\bar{u}^*_\omega ik\hat{A}_{k,n+1} = 0, \tag{20}$$

$$
\begin{aligned}
&A_b(\mu + i(n+\alpha)\omega)\hat{u}_{k,n} + \bar{u}_\omega \frac{i\omega}{2}\hat{A}_{k,n-1} - \bar{u}^*_\omega \frac{i\omega}{2}\hat{A}_{k,n+1} \\
&+u_m A_b ik\hat{u}_{k,n} + \frac{1}{2}\bar{u}_\omega A_b ik\hat{u}_{k,n-1} + \frac{1}{2}\bar{u}^*_\omega A_b ik\hat{u}_{k,n+1} = \\
&-A_b ik\frac{K_e}{\rho}\frac{\hat{A}_{k,n}}{A_o} - \beta_b\pi\nu\hat{u}_{k,n} + \phi_m\hat{A}_{k,n} + \frac{\bar{\phi}_\omega}{2}\hat{A}_{k,n-1} + \frac{\bar{\phi}_\omega}{2}\hat{A}_{k,n+1},
\end{aligned} \tag{21}
$$

2.1.5. *Dimensionless groups.* To simplify the representation, we nondimensionalize the problem via Table S1. Using the dimensionless groups introduced, the nondimensional forms of the characteristic equations 20 and 21 are

**mass conservation equation:**

$$
\begin{aligned}
&(\tilde{\mu} + i(n+\alpha)\tilde{\omega})\tilde{A}_{k,n} + ik''u''_{k,n} + \frac{1}{2}N_m ik''\tilde{A}_{k,n} + \frac{1}{ph[\beta_b]}\frac{N_\omega}{2(2+i\tilde{\omega}/(ph[\beta_b]))}ik''\tilde{A}_{k,n-1} \\
&\qquad + \frac{1}{ph[\beta_b]}\frac{N_\omega}{2(2-i\tilde{\omega}/(ph[\beta_b]))}ik''\tilde{A}_{k,n+1} = 0,
\end{aligned} \tag{22}
$$

**momentum conservation equation:**

$$
\begin{aligned}
&(\tilde{\mu} + i(n+\alpha)\tilde{\omega})u''_{k,n} + \frac{1}{ph[\beta_b]}\frac{N_\omega}{2(2+i\tilde{\omega}/(ph[\beta_b]))}i\tilde{\omega}\tilde{A}_{k,n-1} - \frac{1}{ph[\beta_b]}\frac{N_\omega}{2(2-i\tilde{\omega}/(ph[\beta_b]))}i\tilde{\omega}\tilde{A}_{k,n+1} \\
&+\frac{N_m}{2}ik''u''_{k,n} + \frac{1}{ph[\beta_b]}\frac{N_\omega}{2(2+i\tilde{\omega}/(ph[\beta_b]))}ik''u''_{k,n-1} + \frac{1}{ph[\beta_b]}\frac{N_\omega}{2(2-i\tilde{\omega}/(ph[\beta_b]))}ik''u''_{k,n+1} = \\
&\qquad -ik''\tilde{A}_{k,n} - 2ph[\beta_b]u''_{k,n} + N_m\tilde{A}_{k,n}\frac{\beta_m}{|\beta_b|} + \frac{N_\omega}{2}\tilde{A}_{k,n-1} + \frac{N_\omega}{2}\tilde{A}_{k,n+1},
\end{aligned} \tag{23}
$$

where $\beta_b$ is the complex wall shear coefficient as defined earlier (eqn. 9), and $ph[\beta_b] = \frac{\beta_b}{|\beta_b|}$ has only the phase information $\beta_b$ due to the normalization by the scalar amplitude $|\beta_b|$. Finally, $\tilde{\omega}$ is the dimensionless angular frequency of the cardiac cycle. The important parameters describing the oscillatory component of flow through the blood vessel — including wall shear coefficient $\beta_b$, vessel area $A_m$, pressure driven acceleration $\bar{\phi}_\omega$, and wall stiffness $K_e$ — have been collected in a single dimensionless number

$$
N_\omega = \frac{\bar{\phi}_\omega A_m^{1/2}}{(\frac{|\beta_b|}{2}\pi\nu)}\sqrt{\frac{\rho A_o}{K_e}}. \tag{24}
$$

Akin to the role of the Reynolds number in describing the onset of turbulence, this dimensionless number $N_\omega$ tracks the inception of the flutter type instability at given values of the remaining variables. The other nondimensional number $N_m$ has similar definition as $N_\omega$ (see Table S1) with $\phi_m$ replacing $\bar{\phi}_\omega$. $N_m$ encapsulates the effect of mean flow and its value is typically smaller $(0.05-0.7)$ compared to the values of $N_\omega$ $(0.5-12)$ for physiologic conditions. The dimensionless angular frequency $\tilde{\omega}$ takes values in the range of $12-34$. Finally, the Womersley number $w_0$ has values in the range of $13-35$.

The marginal stability curve $\tilde{\mu}=0$ marks the critical point above which $\tilde{\mu}>0$ perturbation amplitudes grow exponentially in time, and below which $\tilde{\mu}<0$ the base flow is stable under the decay of perturbation modes. To find the locus of points where $\tilde{\mu}=0$, we refer to the methodology proposed by Kumar et al[29]. That is, by fixing the values of $k''$, $\tilde{\omega}$, $N_m$, and $w_0$ (and therefore of $\beta_b$) for a specific flow scenario, as well as presetting $\tilde{\mu}=0$ in eq. 22 and 23, we solve an eigenvalue problem for the critical $N_{\omega,\mathrm{crit}}$ on the marginal stability curve. This procedure is detailed in Supplementary S1.13.

In Fig. 2, we plot the harmonic $\alpha = 1$ and subharmonic $\alpha = 1/2$ "tongues" of instability[29]. The space of dimensionless wavenumber $k''$ and dimensionless number $N_\omega$ is divided into tongue regions of instability, where perturbations to the flow grow in time, and outside zones of stability, where the base flow remains stable to perturbations. The harmonic response has the same frequency $\tilde{\omega}$ (and its multiples) as that of the driving pressure gradient in the base flow whereas the subharmonic response has half the frequency $\tilde{\omega}$ (and its multiples). The subharmonic solution is excited first as $N_\omega$ increases past the lowest threshold value $N_{\omega,\text{threshold}}$. This critical threshold value occurs at the bottom tip of the first subharmonic tongue. It is the global minimum of the critical dimensionless number on all tongues of marginal stability, $\min(N_{\omega,crit}) = N_{\omega,\text{threshold}}$ (see Fig. 2).

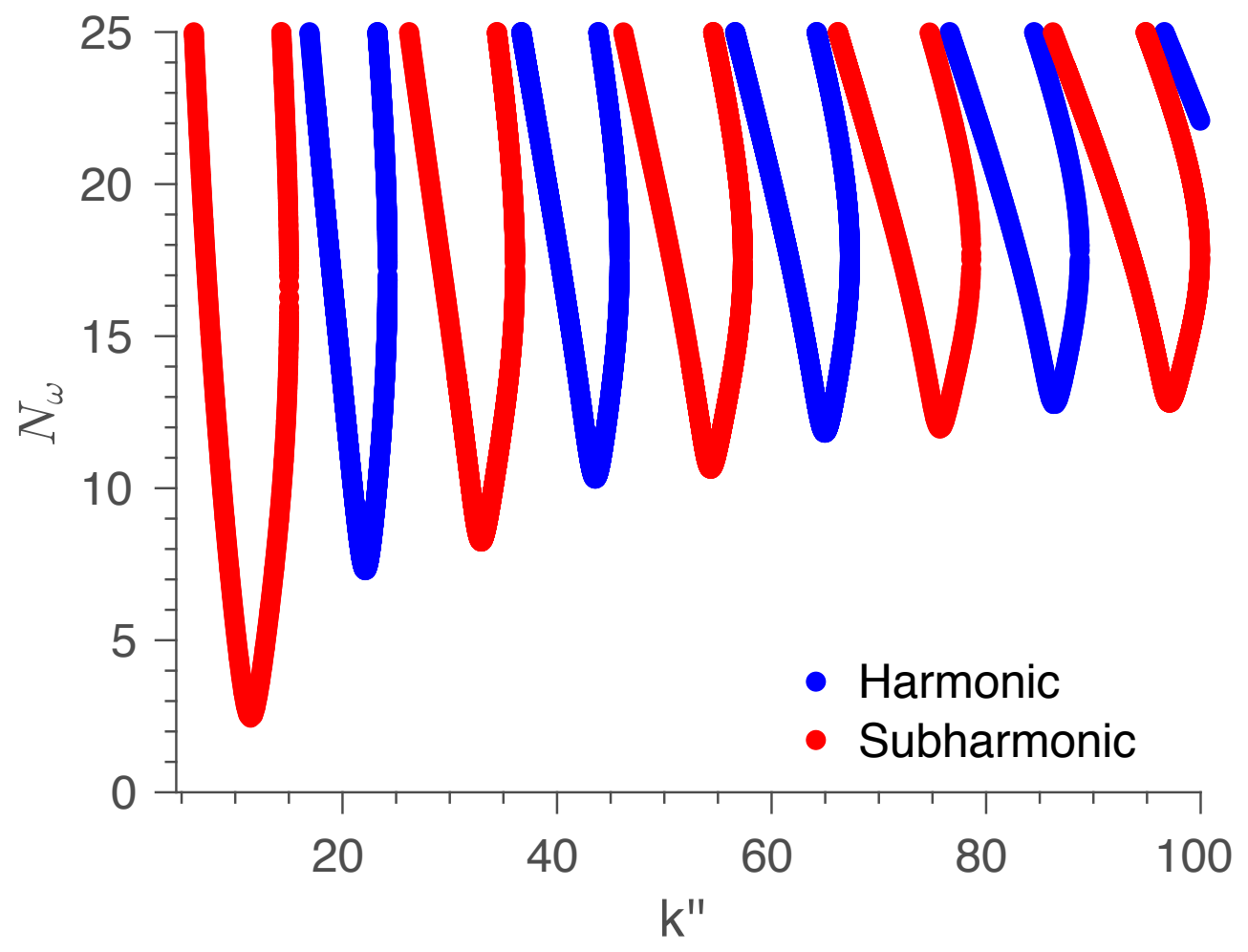

FIGURE 2. The marginal stability curve $\tilde{\mu} = 0$ as a function of the dimensionless wave number $k''$ of the perturbation mode and the dimensionless number $N_\omega$. The dimensionless number $N_\omega$ encapsulates the blood viscosity, vessel diameter, pressure gradient (or flow acceleration), and viscous contribution under pulsatile waveform of the flow. For a specific value of $k''$, $N_m$, and $\tilde{\omega}$, $N_\omega$ within the alternating tongues indicate that the system is unstable to perturbations and can grow unboundedly, whereas $N_\omega$ outside the tongues correspond to stable base flow. The figure uses representative values of the angular frequency $\tilde{\omega} = 19.6$ and $N_m = 1.7 \times 10^{-1}$ corresponding to human physiology.

If $N_\omega > N_{\omega,\text{threshold}}$, the blood vessel will be unstable to a waveband of perturbation modes, whereas below this threshold, the base flow should remain stable. We hypothesize that the growth of perturbation modes will trigger or signal the permanent dilatation of a cross-sectional area of the blood vessel over time. We may test whether the dimensionless number is predictive of future aortic growth and potential aneurysm development by measuring the patient specific physiological properties comprising $N_\omega$ (e.g., through cardiac imaging) and validating this theoretical forecast against observed aortic dilatation at follow-up.

2.2. **Pulse wave velocity.** To determine the flow stability for a specific patient, the above formulation requires information about the wall stiffness $K_e$ of the blood vessel. This physiological property can be found from the pulse wave velocity (PWV) measured from imaging techniques such as MRI scans and echocardiograms. The PWV is the propagation speed of the pulse wave in the aorta and is related to the elastic modulus or stiffness of the aortic wall. This relationship can be derived by transforming the set of simplified governing equations to the standard form of the wave equation[30]. The precise steps are listed in Supplementary S1.2.

Although a linear tube law was used in the derivation of the dimensionless number, a general tube law is likewise permissible via $P = \frac{1}{\rho}G(A)$, where $G$ can be some nonlinear function of the local cross-sectional area. The function $G$ represents the full dependence of the excess internal pressure on the cross-sectional area and thus can encapsulate aortic wall properties such as elastic moduli, wall thickness, etc. in the most general case. The pulse wave velocity is then shown in Supplementary S1.2 to be

$$c_{pw}^2 = \frac{1}{\rho}\frac{\mathrm{d}G}{\mathrm{d}A}A. \tag{25}$$

In the context of the dimensionless number derivation, the tube law term appears in the linearized perturbation equations. By expansion around the base pressure $P_b$ and area $A_m$,

$$P = P_b + P' = P_b + \frac{1}{\rho}\frac{\mathrm{d}G}{\mathrm{d}A}\Big|_b A' = P_b + c_{pw}^2\frac{A'}{A_m}. \tag{26}$$

We see that no matter which form the tube law $G(A)$ takes, the measured PWV can be used to quantify the blood vessel's elastic properties. The key dimensionless number can be recast in terms of PWV as follows

$$N_\omega = \frac{\bar{\phi}_\omega A_m^{1/2}}{(\frac{|\beta_b|}{2}\pi\nu)}\sqrt{\frac{\rho A_o}{K_e}} = \frac{\bar{\phi}_\omega A_m}{\frac{|\beta_b|}{2}\pi\nu c_{pw}}. \tag{27}$$

Using eqn. 27, $N_\omega$ can now be calculated explicitly from clinical imaging data for each cross-section along a blood vessel. The difference between this clinical patient specific value $N_\omega$ and the critical threshold $N_{\omega,\mathrm{threshold}}$ on the marginal stability curve produces an overall flutter instability parameter

$$N_{\omega,\mathrm{sp}}[N_m, \tilde{\omega}, \beta_b[w_0]] = N_\omega - N_{\omega,\mathrm{threshold}}[N_m, \tilde{\omega}, \beta_b[w_0]]. \tag{28}$$

Eqns. 22 and 23 imply that $N_{\omega,\mathrm{threshold}}$ depends on $N_m, \tilde{\omega}$, and $\beta_b$. This is reflected in eqn. 28 from the functional dependence of $N_{\omega,\mathrm{threshold}}$ and by consequence of $N_{\omega,\mathrm{sp}}$ on these variables. Note that $\beta_b$ depends on the Womersley number $w_0$ (eqn. 9). All the independent variables $(N_m, \tilde{\omega}, w_0)$ can be determined clinically.

If $N_{\omega,\mathrm{sp}} > 0$, we hypothesize that the blood vessel cross-section is expected to grow due to the increase in perturbation amplitude. Otherwise for $N_{\omega,\mathrm{sp}} \leq 0$, the blood vessel diameter should remain constant in time since all perturbation modes decay. Thus, the flutter instability parameter $N_{\omega,\mathrm{sp}}$ can serve as an aneurysm physiomarker that is predictive of abnormal aortic growth and is convenient to apply clinically.

In summary, we have developed an ab initio theoretical framework to predict the stability of an aortic section depending on a patient's aorta diameter $A_m$, blood pressure gradient $\bar{\phi}_\omega$ causing oscillatory acceleration, pulsatile contribution to wall shear $\beta_b$, blood viscosity $\nu$, and blood density $\rho$. These values can be extracted from 4D flow MRI or reference literature[31].

## 3. Clinical application

3.1. **Study cohorts.** To gauge the performance of the proposed aneurysm physiomarker in analyzing aneurysm growth, a retrospective study was carried out for subjects with and without existing aortopathies.

3.1.1. *Overall patient cohort.* Patients were respectively identified from a database of patients who underwent a clinical cardiothoracic MRI exam, including 4D flow MRI, at Northwestern Memorial Hospital between 2011 and 2019. Inclusion criteria were referral for clinical imaging assessment of aortic dimensions and a normal tricuspid aortic valve (TAV). Exclusion criteria were presence of aortic valve stenosis (mild to severe), ejection fraction lower than 50%, or bicuspid aortic valve. In addition, patients with 4D flow MRI data that have not undergone dedicated analysis (eddy current and concomitant phase corrections, aortic 3D segmentation) were excluded. A summary of the selection process with inclusion and exclusion counts is given in Fig. S2.

A total of 125 patients were identified for inclusion in this study. Of these, 8 patients were excluded due to 4D flow MRI imaging artifacts, resulting in 117 patients for this study. All patients in this HIPAA compliant study were retrospectively included with approval from the Northwestern University Institutional Review Board (IRB) and IRB-approved waiver of consent. Records were de-identified prior to analysis.

3.1.2. *Healthy subject cohort.* For comparison to the overall patient cohort, a total of 100 healthy control subjects were included, evenly distributed across a wide range of ages and sexes (age range 19 years to 79 years, 50% female). The 100 healthy subjects included were selected from a group of healthy subjects who had been prospectively enrolled for research MRI exams under a separate IRB-approved protocol. Informed consent was provided by all study participants. The group of 100 healthy control subjects selected for analysis was created by taking the first ten (chronological order) recruited subjects of each sex, divided into the age ranges of 19-30, 31-40, 41-50, 51-60, and 61-79 years. The control cohort demographics are summarized in Table 3.

3.1.3. *Subcohort for patient outcomes classification.* To evaluate the predictive performance of $N_{\omega,\mathrm{sp}}$, a subcohort (labeled prognosis aortopathy patients) was created for patients with follow-up measurement of aortic dimensions. The inclusion criterion was having magnetic resonance angiography (MRA) or computed tomography angiography (CTA) aortic dimensions assessment within five years of initial 4D flow imaging. Exclusion criteria for this subcohort were presence of genetic tissue disorders, congenital heart malformations, and history of aortic or mitral valve repair occurring before the 4D flow imaging analyzed (Fig. S2). Of the 117 patients, 25 patients lacked follow-up imaging, 3 had a history of dissection, 14 history of aortic repair, and 3 had Marfan syndrome. The final prognosis patient subcohort included 72 patients.

In this subcohort of patients, two outcomes were quantified: "growth" and "surgery". The aortic diameter growth was assessed from radiological measurements taken with CT or MR angiography imaging, which included standardized assessment of the maximal-area ascending aorta (MAA) and the sinus of valsalva (SOV) diameters in double-oblique view. During the follow-up period, any intervention, such as valve repair or aortic graft placement, that occurred after the 4D flow imaging was used to categorize patients as having "surgery" outcomes. Additional details of image acquisition and processing are described in Supplementary S1.5 and S1.6, respectively.

As an illustration of how growth outcomes are calculated, the maximum of the SOV and MAA diameters recorded during each clinic visit ($\mathrm{SOV}_{\max}$ and $\mathrm{MAA}_{\max}$ ) are presented in two time series after the initial MRI at year 0 (Fig. S4**A**, S4**B**). The growth rate was then calculated as the maximum rate of change over time between consecutive pairs of follow-up assessments. That is, the maximum SOV growth rate was characterized as $\Delta\mathrm{SOV}_{\max} = \max_{\forall t}(\frac{d\mathrm{SOV}_{\max}}{dt})$, and the maximum MAA growth is characterized as $\Delta\mathrm{MAA}_{\max} = \max_{\forall t}(\frac{d\mathrm{MAA}_{\max}}{dt})$. A diameter change in SOV or MAA of 0.24 cm/year or greater was then categorized as an abnormal "growth" outcome for the patient.

An example of a patient's time-series is shown in Fig. S4**A**, where $\Delta\mathrm{SOV}_{\max} = 0.05$ cm/year due to a stepwise jump in measured $\mathrm{SOV}_{\max}$ between years 2 to 3. This growth rate is defined analogously for the maximum MAA diameter; Fig. S4**B** gives $\Delta\mathrm{MAA}_{\max} = 0.14$ cm/year for the same patient. Since $\Delta\mathrm{MAA}_{\max} < 0.24$ cm/year, $\Delta\mathrm{SOV}_{\max} < 0.24$ cm/year, and this patient also did not undergo any surgery during follow-up, the classification "no growth or surgery" was applied.

## 4. Results

4.1. **Patient aortic growth and $N_{\omega,\mathrm{sp}}$ predictive performance.** Since the flow velocities are resolved spatially through 4D-flow MRI, the aneurysm physiomarker $N_{\omega,\mathrm{sp}}$ can be visualized based on location along the centerline of the aorta as a result of our 1D analysis (Fig. 3). The aneurysm physiomarker is calculated from the patient's initial MRI taken at year 0 and can be evaluated against follow-up data that report SOV and MAA diameters. For instance, Fig. 3**A** shows that a patient's SOV and MAA growth rates of 0.38 cm/year and 0.15 cm/year agree with their spatial aneurysm physiomarker distribution; $N_{\omega,\mathrm{sp}} > 0$ is localized near the SOV rather than the MAA. Similarly, Fig. 3**B** demonstrates that a second patient's SOV and MAA growth rates of 0.08 cm/year and 0.30 cm/year likewise matches their aneurysm physiomarker distribution; here, $N_{\omega,\mathrm{sp}} > 0$ occurs at the MAA rather than the SOV. The growth rates in SOV and MAA vs their respective, locally measured aneurysm physiomarker are show in Fig. S6.

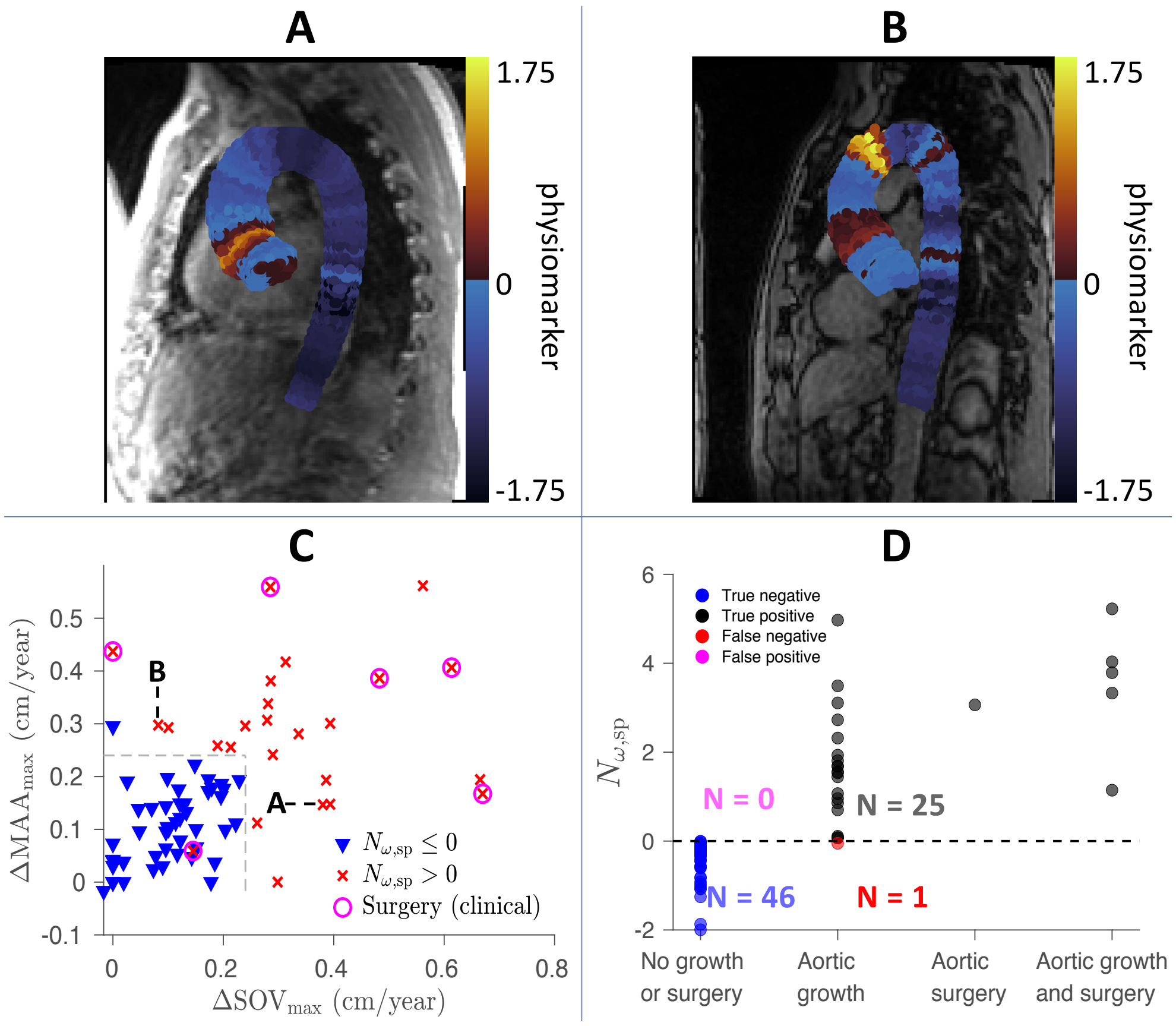

FIGURE 3. **A**) Example of the 1D spatial distribution of the flutter aneurysm physiomarker $N_{\omega,\text{sp}}$ along the central axis for one patient, calculated from their initial MRI taken at year 0. Each slice perpendicular to the centerline is colored the same to represent the value along the central axis. During clinical follow-up, the patient exhibited growth rates of 0.38 cm/year and 0.15 cm/year at their sinus of valsalva (SOV) and maximal-area ascending aorta (MAA). This agrees with the aneurysm physiomarker distribution, which shows $N_{\omega,\text{sp}} > 0$ localized near the SOV rather than the MAA. **B**) Example of the spatial distribution of the aneurysm physiomarker for a second patient, who exhibited growth rates of 0.08 cm/year & 0.30 cm/year at the SOV & MAA. These rates likewise match the aneurysm physiomarker distribution, where $N_{\omega,\text{sp}} > 0$ at the MAA rather than the SOV. **C**) A prediction vs outcome diagram of all patients with follow-up imaging data. The maximum growth rate of their MAA and SOV (cm/year) measured from follow-up imaging data are visualized with respect to the theoretical prediction $N_{\omega,\text{sp}}$, which are measured from a single MRI at time 0. If $N_{\omega,\text{sp}} > 0$, the patient's marker is labeled by an x. Otherwise, the data point is labeled by a downward pointing triangle. The circles indicate that the patient experienced a surgical intervention after their initial MRI at year 0. The growth boundary of 0.24 cm/year is labeled by black dotted lines. This boundary optimally discriminates between stable and unstable aneurysms predicted by the proposed aneurysm physiomarker and falls within the clinically observed range of abnormal growth (0.24 cm/year for small aneurysms to 0.31 cm/year for large aneurysms) that is associated with chronic dissection [7]. The two patients imaged in **A** and **B** are marked appropriately. **D**) Each patient has been labeled according to whether $N_{\omega,\text{sp}} > 0$ accurately predicts a growth outcome (categorized as exhibiting a growth rate in SOV or MAA $\geq 0.24$ cm/year), a surgical intervention, or both at follow-up.

Next, the per patient growth rates for the SOV and MAA are visualized in Fig. 3**C** and compared with our theoretical predictions. Each "$\times$" in Fig. 3**C** denotes $N_{\omega,\mathrm{sp}} > 0$, as calculated from the patient's MRI image at year 0. This indicates that the ascending aorta is expected to grow due to the flutter type instability. Conversely, each "$\nabla$" represents $N_{\omega,\mathrm{sp}} \leq 0$. Since all perturbation modes are damped in this case, the ascending aorta should not be subject to the identified instability. Data points for patients who experienced surgical intervention after their initial MRI are circled. All aneurysm physiomarker values $N_{\omega,\mathrm{sp}}$ were calculated from patient MRI at time zero, without reference to follow-up data.

Growth rates exceeding 0.2 cm/year lie outside the range of normal growth of the thoracic aorta[7]. When the stability of aneurysm measured at time zero via $N_{\omega,\mathrm{sp}}$ are plotted with respect to the growth rates measured from follow-up data, we find that a growth threshold of 0.24 cm/year optimally discriminates between stable and unstable aneurysms forecast by the proposed aneurysm physiomarker. This is an emergent division of the growth data based purely on the transition of $N_{\omega,\mathrm{sp}}$ from negative to positive- from natural aortic dilatation over time to abnormal growth driven by unstable flutter.

The proposed discriminating boundary of 0.24 cm/year falls within the clinically observed range of significantly higher growth rates (0.24 cm/year for smaller 4 cm aneurysms to 0.31 cm/year or larger 5.2 cm aneurysms) that is associated with chronic dissection in patients[7]. Growth between 0.2 to 0.3 cm/year thus appears to delineate a transition zone from low to high risk. For instance, the growth rate in a non-referral, low risk population of patients with ascending thoracic aneurysms (> 4 cm) ranged from 0.07 to 0.16 cm/year[32]; meanwhile, European Society of Cardiology (ESC) guidelines[33] have suggested that growth exceeding 0.3 cm/year is a risk factor that can prompt surgical intervention for thoracic ascending aortic aneurysms.

This agreement between the emergent boundary based on the flutter aneurysm physiomarker and the statistically significant growth rate that clinicians have independently deduced validates the aneurysm physiomarker as an unbiased, clinically valuable predictor of abnormal aneurysm growth. This boundary for abnormal growth has been visualized in Fig. 3**C**.

Fig. 3**D** shows that by using the aortic growth rate of 0.24 cm/year as an indicator of significant growth, the aneurysm physiomarker $N_{\omega,\mathrm{sp}} > 0$ serves as a good binary predictor for the growth outcome of each patient. The accuracy, sensitivity, and specificity of this proposed aneurysm physiomarker in predicting abnormal growth in the thoracic aorta are 0.986, 0.962, and 1.000, respectively. The area under the curve (AUC) of a receiver operating characteristic (ROC) analysis (Fig. S7) is 0.997; an AUC above 0.9 is typically considered "outstanding" for the performance of a binary predictive diagnostic[34].

Additionally, the optimal operating point occurs at the minimum positive value for $N_{\omega,\mathrm{sp}}$ for patients with follow-up data, suggesting that the analytically derived threshold $N_{\omega,\mathrm{threshold}}$ accurately describes the onset of the underlying instability. No training data set was necessary to tune the calculation of the aneurysm physiomarker for each patient. If the more conservative threshold 0.31 cm/year is selected instead as a binary indicator of clinically significant growth, the accuracy, sensitivity, specificity, and AUC of this proposed aneurysm physiomarker become 0.875, 1.000, 0.8393, and 0.952 respectively. Its performance in classifying abnormal growth is therefore bounded from below in the "outstanding" category for typical clinical use cases[34].

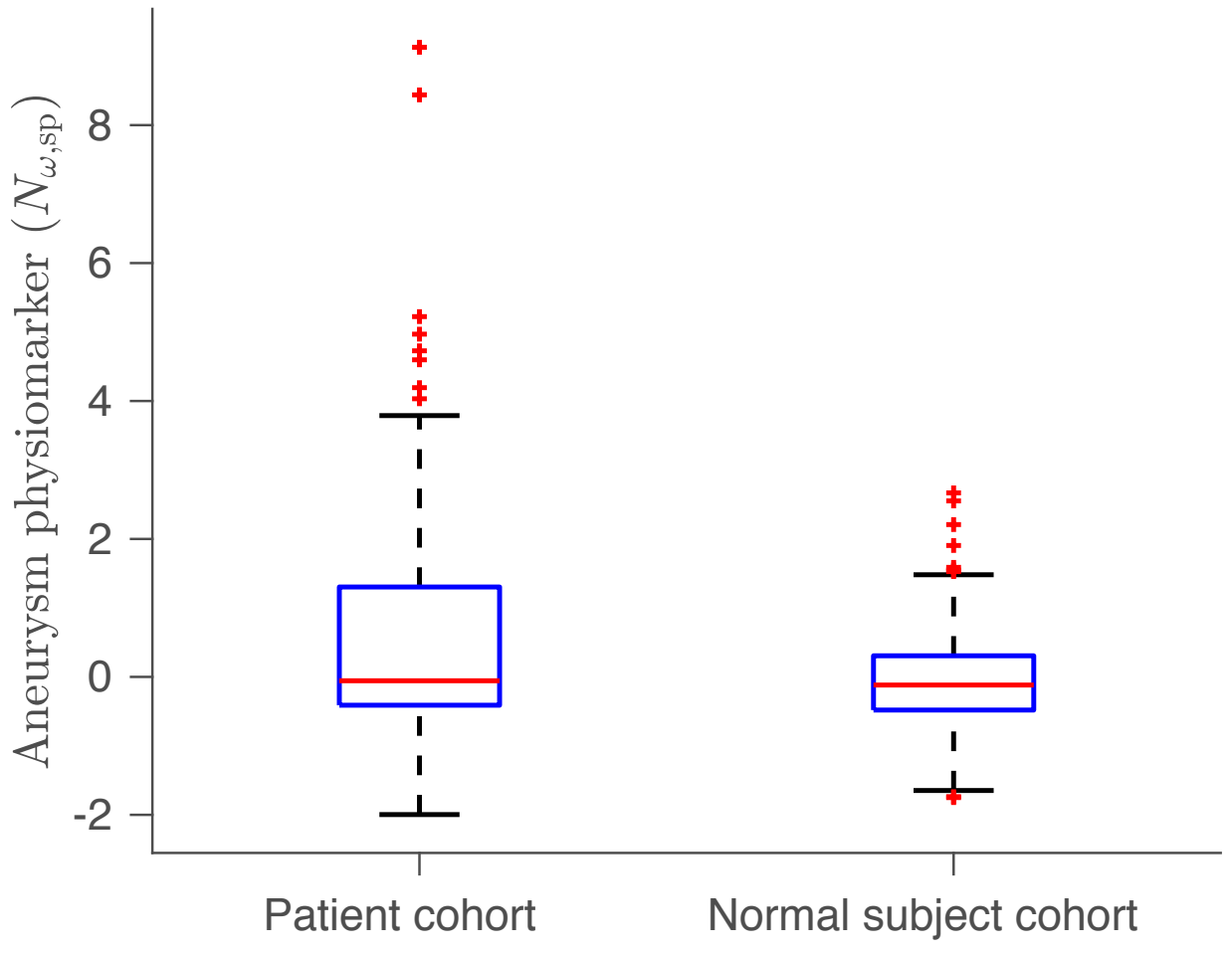

FIGURE 4. **A**) The distribution of the aneurysm physiomarker $N_{\omega,\mathrm{sp}}$ in the patient and normal subject cohorts. The median aneurysm physiomarker value for the normal subject cohort is shown to be significantly ($p<0.05$) smaller than that for the patient cohort, via a one-tailed Wilcoxon rank sum test. Patient cohort: {min, 25% quartile, median 75% quartile, max} = {-1.9961, -0.4107, -0.0561, 1.3035, 9.1285}. Normal subject cohort: {min, 25% quartile, median 75% quartile, max} = {-1.7499, -0.4807, -0.1180, 0.3061, 2.6680}

4.2. **Cohort comparisons.** Next, the distributions of the aneurysm physiomarker $N_{\omega,\mathrm{sp}}$ in both the normal subject cohort and the patient cohort are examined. As seen in Fig. 4, the median aneurysm physiomarker value for the normal subject cohort is shown to be significantly ($p= 0.0370$) smaller than that for the patient cohort, via a one-tailed Wilcoxon rank sum test. This agrees with the inclusion criteria used to establish the patient and normal subject cohorts; Fig. S4 shows that the aneurysm physiomarker $N_\omega$ appears to trend with increased growth rates in the SOV and MAA. The sample size of both cohorts exceed 93, the value required to establish significance at a level of $p<0.05$ for the difference in their median values with 90% statistical power.

To neutralize potential bias in the statistics arising from lack of age or sex matching, we also binned the aneurysm physiomarker measured for patient and normal subject cohorts into different age and sex groups in Table S2. The female normal subjects in the youngest age group (Age $<$ 40) show a significantly smaller aneurysm physiomarker value compared to males in the same cohort. This could reflect population level observations that TAAs occur more commonly in males than in females, despite poorer outcomes in females[35]. We also note that for the patient cohort, females exhibit systematically though not significantly higher $N_{\omega,\mathrm{sp}}$ than males across every age group. This sex disparity may mirror clinical observation that TAA growth is accelerated in females compared to males[35]. Thus, the distribution of the aneurysm physiomarker among different sex and age groups in the two cohorts appear to agree with general population trends reported in the literature.

## 5. Discussion

5.1. **Predictive power of the aneurysm physiomarker.** While the aneurysm physiomarker proves predictive of abnormal growth, we do not expect it to discriminate between no growth and any nonzero amount of aortic dilatation. After all, increase in aortic dimensions occurs with natural aging. Normal aortic growth, while complex in etiology, is generally understood to occur as a result of the repeated natural stresses induced by pressurized blood flow, which ultimately result in gradual loss of elastin fibers, remodeling of elastic lamellae, and ultimately increase of vessel diameter[36;37]. Normative rates of increase in adults have been assessed at 0.11 cm/year in adults for both men and women, but several factors such as blood pressure and body surface area are associated with larger diameters. There is high variation in baseline aortic diameter at any given age range[38;39;37].

Instead of describing all modes of aortic dilatation, the aneurysm physiomarker identifies the specific presence of the flutter instability, which appears to signal subsequent abnormal growth for a significant percentage of patients who eventually experience rates exceeding 0.24 cm/year. Thus, the aneurysm physiomarker's ability to predict abnormal dilatation in contrast to natural growth is crucial for prompt clinical decision-making and accurate treatment[7].

The aneurysm physiomarker may also expand aneurysm detection and prediction to aortic segments heretofore less commonly examined due to the effort and cost involved. Interestingly, the spatially resolved aneurysm physiomarker distribution in Fig. 3**B** displays a global maximum near the aortic arch as well as multiple pockets of instability $N_{\omega,\mathrm{sp}} > 0$ along the descending aorta; this spatial variation in the aneurysm physiomarker is largely due to local change in the pressure driven acceleration $\bar{\phi}_\omega$ and area $A_b$. This demonstrates that abnormal aortic dilatation is not necessarily confined near the SOV and MAA. More comprehensive imaging analysis motivated by predictive aneurysm physiomarker distributions may help detect "silent-until-rupture" aneurysms that evade screening at common sites[40]. It would be of strong interest to conduct a complete chart review showing whether such spatial aneurysm physiomarker predictions can be confirmed by follow-up imaging and clinical intervention; this procedure is out of the scope of the current work.

5.2. **Aneurysm development in normal subjects with respect to age.** Although age is not a direct input into the eigenvalue analysis that yields the aneurysm physiomarker $N_{\omega,\mathrm{sp}}$, many physiological properties vary systematically with age. For instance, both aortic diameter and wall stiffness are known to increase naturally in older, healthy subjects. Less is known about how these age related variations affect aneurysm formation and growth. We elucidate important trends here.

Table 4 shows the breakdown of physiological properties that make up the aneurysm physiomarker $N_{\omega,\mathrm{sp}}$. This comparison occurs across three age groups. In the normal subject cohort, an unstable flow condition $N_{\omega,\mathrm{sp}} > 0$ is induced on average by two significant factors relative to the stable $N_{\omega,\mathrm{sp}} \leq 0$ normal subjects − the larger pressure gradient $\bar{\phi}_\omega$ that causes blood flow oscillatory acceleration as well as the smaller pulse wave velocity $c_{pw}$. $N_{\omega,\mathrm{sp}}$ becomes positive primarily because $N_{\omega,\mathrm{clin}}$ (eqn. 28) increases for larger $\bar{\phi}_\omega$ and smaller $c_{pw}$, while $N_{\omega,\mathrm{threshold}}$ decreases for larger $\bar{\phi}_\omega$ and smaller $c_{pw}$.

For the youngest age group (Age $<$ 40), the dominant factor is larger (p = 0.0081) pressure gradient $\bar{\phi}_\omega$. That is, normal subjects with unstable aneurysm physiomarker $N_{\omega,\mathrm{sp}} > 0$ exhibit larger $\bar{\phi}_\omega$ compared to normal subjects with a stable aneurysm physiomarker $N_{\omega,\mathrm{sp}} \leq 0$ in a one-tailed Wilcoxon rank sum test. The role of greater blood flow acceleration is well established as a qualitative marker in aneurysm development; hypertension is well acknowledged as a risk factor in aneurysm formation and growth[41] and has been implicated in modulating the morphology of unstable aneurysms[42].

In the middle age group (40 $\leq$ Age $<$ 60), the pressure gradient $\bar{\phi}_\omega$ is likewise significantly higher for unstable aortic flow $N_{\omega,\mathrm{sp}} > 0$. The second factor that appears is smaller (p = 0.029) pulse wave velocity $c_{pw}$, which indicates lower aortic stiffness. Compliant aortic walls distend farther and can sustain more unstable flutter modes under the same pressure gradient compared to stiffer aortas characterized by higher $c_{pw}$. Thus, the natural stiffening of the aorta with age for healthy subjects serves to protect against further dilatation. This explains why the oldest age group (Age $\geq$ 60) has no normal subjects exhibiting $N_{\omega,\mathrm{sp}} > 0$.

5.3. **Aneurysm development in patients with respect to age.** In the patient cohort (Table 4), flutter instability aneurysm physiomarker $N_{\omega,\mathrm{sp}} > 0$ is mainly driven by a smaller pulse wave velocity $c_{pw}$ compared to patients experiencing stable flow $N_{\omega,\mathrm{sp}} \leq 0$. In the youngest age group (Age $< 40$), we observe that aortic area $A_m$ is likewise significantly higher for patients with positive aneurysm physiomarker values. This matches clinical observations of increased dilatation risk with larger aneurysm size[6]. As before, $N_{\omega,\mathrm{sp}}$ becomes positive primarily because $N_{\omega,\mathrm{clin}}$ (eqn. 28) increases and $N_{\omega,\mathrm{threshold}}$ decreases for smaller $c_{pw}$. Larger aortic area also leads to increased $N_{\omega,\mathrm{clin}}$ and, to a lesser extent, increased $N_{\omega,\mathrm{threshold}}$, such that the aneurysm physiomarker (eqn. 28) increases overall.

In every age group for the patient cohort, the median pulse wave velocity is significantly lower for patients with unstable aortic flows $N_{\omega,\mathrm{sp}} > 0$ compared to stable patients $N_{\omega,\mathrm{sp}} \leq 0$. This suggests that greater wall distensibility plays a dominant role in facilitating growth of larger, developed aneurysms in the patient cohort. Permanent dilatation occurs when the aortic wall weakens and becomes less stiff. Such a process can form a self-perpetuating cycle, since thinning of the intimal and medial layers during aneurysm expansion increases aortic distensibility, which supports further dilation by increasing aortic wall susceptibility to unstable flutter modes.

A summary of clinical observations on how aneurysm distensibility evolves during disease progression is provided in Table S3. As noted earlier, the aortic wall degrades due to elastin and smooth muscle loss through aneurysm enlargement[9;10;11]. Hereafter, collagen deposition either stiffens the aortic wall (no further growth; Type 1), or the aortic wall weakens due to lack of collagen deposition, wall inflammation, and/or adipocyte accumulation (Type 2)[12]. The latter can lead to eventual dissection or rupture[11].

Our proposed aneurysm physiomarker provides support for these clinically observed pathways. Patients who exhibit stable flows $N_{\omega,\mathrm{sp}} \leq 0$ have significantly larger pulse wave velocities and therefore fall within the Type 1 "stiff" aneurysm group. On the other hand, every patient age group with unstable aortic flows $N_{\omega,\mathrm{sp}} > 0$ has significantly lower pulse wave velocity than stable patients in the same age group. Thus, patients whose compliant aortic walls fail to respond and lay down collagen remain vulnerable to growth driven by the flutter instability. This indicates that unstable patients possess Type 2 "soft/at-risk" aneurysms.

5.4. **Cross cohort comparisons of aneurysm drivers.** Next, we compare different physiological properties driving aneurysm growth between the normal subject cohort and the patient cohort.

In the youngest age group (Age $< 40$), the stable $N_{\omega,\mathrm{sp}} \leq 0$ patient cohort exhibits a significantly larger median pulse wave velocity than stable normal subjects. This further reinforces the clinical observation that the branch of aneurysm progression toward the stable Type 1 aneurysm is marked by stiffening of the aortic wall that prevents additional dilatation. Similarly, the unstable $N_{\omega,\mathrm{sp}} > 0$ patient cohort exhibits a significantly smaller median pulse wave velocity than stable (p=$5 \times 10^{-4}$) normal subjects. Thus, unstable patients comprise the second trajectory of aneurysm development $-$ the Type 2 aneurysm group for which increased wall distensibility triggers further growth.

As the age of normal subjects increases through the three defined groups, we observe that the pulse wave velocity of stable normal subjects $N_{\omega,\mathrm{sp}} \leq 0$ increases significantly from (Age $< 40$) to ($40 \geq$ Age $< 60$) with (p $=0.0011$), as well as from ($40 \geq$ Age $< 60$) to (Age $\geq 60$) with (p $= 0.0011$). This reflects the natural stiffening of the aorta with age and also serves to protect against unstable flutter. However, we note that the median pulse wave velocity of the youngest (Age $< 40$) stable normal subject cohort is still significantly larger than that of unstable patients of any age group. The Type 2 progression of aortic aneurysms therefore marks a diseased state in which distensibility increases abnormally above reference, healthy values due to wall remodeling. This disease trajectory is especially prominent in the oldest age group (Age $\geq 60$), where the pulse wave velocity for the unstable patient cohort is significantly lower than that of the stable normal subject cohort (p $=5 \times 10^{-7}$). Natural stiffening of the aorta has entirely failed to kick in for the unstable patient cohort and is replaced by aneurysmal weakening of the wall. Thus, the trends obtained for our ab initio aneurysm physiomarker are in good agreement with observed tissue biology during aneurysm development and it provides a quantitative, noninvasive prediction of anticipated growth.

The aneurysm physiomarker trends also elucidate other physiological drivers that contribute to abnormal aortic dilatation. For instance, we observe that the initial growth of aneurysms in normal subjects is driven mainly by a significantly larger pressure gradient $\bar{\phi}_\omega$ for (Age $<$ 40) and (40 $\geq$ Age $<$ 60). Without the associated wall stiffening to constrain these unstable modes, due to a failure to remodel or insufficient response time relative to growth progression, abnormal aortic dilatation occurs. Meanwhile, the abnormal growth of aneurysm in the patient cohort for all age groups is driven primarily by lower pulse wave velocity, as we have already examined in depth. Thus, the fundamental physiology responsible for aneurysm progression varies significantly depending on whether the subject is in an earlier or later stage of the disease. Different treatment options and drug targets would then be necessary to address the root cause of abnormal growth for each patient depending on the dominant physiological property associated with or triggering the flutter instability. Quantitatively, this can be defined by measuring the sensitivity of $N_\omega$ to factors like pressure gradient $\bar{\phi}_\omega$ associated with blood oscillatory acceleration and pulse wave velocity $c_{pw}$. For instance, if reducing $\bar{\phi}_\omega$ to a manageable level would bring the aneurysm physiomarker $N_\omega$ below 0, indicating stable flow, then blood pressure management may be the preferred course of treatment for a patient.

Finally, we note that the median aortic area $A_m$ and oscillatory wall shear coefficient $\beta_b$ are both significantly larger for stable patients compared to stable normal subjects in the age groups (40 $\geq$ Age $<$ 60) to (Age $\geq$ 60). In the same age groups however, these two physiological properties are not significantly different between stable patients and unstable patients, nor for stable normal subjects and unstable normal subjects. This suggests that larger $A_m$ and $\beta_b$ accompany disease progression and may differentiate between subjects who have already experienced aortic dilatation, but not necessarily drive further, abnormal growth on a consistent basis.

In existing literature, many ambiguous observations surround each of the individual physical properties examined in Table 4. For instance, a definite relationship between hypertension and aneurysm growth is not apparent[13], especially since patients without hypertension can likewise experience both aneurysm growth and rupture[42]. High blood pressure is often interpreted as the mechanism driving increased shear stress along the aortic walls, but both high wall shear stress, low wall shear stress, and the spatio-temporal heterogeneity of wall shear stress have been implicated in wall remodeling and aneurysm growth[14;43]. Similarly, larger aortic size is known to correlate with increasing risk of rupture[6], but it is unclear why this is the case.

The aneurysm physiomarker presented in the current work explains not just how these properties trend at the cohort level, but also reveals the mechanism of how they interact explicitly in each patient. For instance, this aneurysm physiomarker framework suggests a future study to clearly delineate the role of shear stress in driving flutter at different parametric conditions. Under varying values of aortic area $A_m$, pulse wave velocity $c_{pw}$, and other physiological variables, wall shear can exhibit nonlinear, non-monotonic dependencies with $N_{\omega,\mathrm{sp}}$ that may account for the significant breadth of prior clinical observations.

The proposed flutter aneurysm physiomarker clarifies the role of each physical property in driving the flutter type instability and delineates the threshold which separates stable aneurysms from unstable growth. These physiological properties cannot be used to predict abnormal dilatation on their own without knowing their relative, quantitative role in driving or inhibiting aneurysm growth for each patient − this, we propose, is the key problem resolved by the aneurysm physiomarker $N_{\omega,\mathrm{sp}} > 0$.

5.5. **Limitations.** While 4D-flow MRI provides a resolved spatial view of flow variables and aneurysm dimensions, it is time limited to a window of one single heartbeat. The physiological variables measured in this interval may not necessarily be representative of a patient's average daily hemodynamic flow conditions. Variability in physiological properties is not described.

5.5.1. *Sensitivity analysis.* Thus, we gauge the local sensitivity of the aneurysm physiomarker toward uncertainty in the measurement of the input physiological properties. This analysis also estimates the error incurred by the constant kinematic viscosity value $\nu$ assumed for every subject; the finite spatial and temporal resolution of 4D flow MRI (S1.5); and retest variation[44]. Table 1 reports the resulting change in $N_{\omega,\text{sp}}$ given an $\epsilon \in [5, 10, 15]\%$ variation of the individual parameters around the measured values for the patient cohort. The aneurysm physiomarker proves most sensitive to the pulse wave velocity $c_{pw}$ and pressure driven acceleration $\bar{\phi}_\omega$, both of which are measured algorithmically from the mean cross-sectional velocity calculated from 4D flow imaging (S1.6). The kinematic viscosity $\nu$ is of tertiary yet nonnegligible importance; a rigorous method to calculate blood viscosity and density through imaging or other non-invasive methods is therefore highly desirable, but out of the scope of the current work.

Table 1. The fundamental physiological properties (e.g. pressure driven acceleration $\bar{\phi}_\omega$) that contribute to evaluating the aneurysm physiomarker $N_{\omega,\text{sp}}$ for 117 patients are varied by a total range of $2\epsilon\%$ around either the measured value or the assumed constant value (kinematic viscosity) in a local sensitivity analysis (e.g. $\bar{\phi}_\omega \pm (\epsilon\%)\bar{\phi}_\omega$). $\epsilon$ is varied from 5 to 15. The magnitude of the resulting change $\Delta N_{\omega,\text{sp}}$ (e.g. $|N_{\omega,\text{sp}}(\bar{\phi}_\omega + (\epsilon\%)\bar{\phi}_\omega) - N_{\omega,\text{sp}}(\bar{\phi}_\omega - (\epsilon\%)\bar{\phi}_\omega)|$) is reported as mean $\pm$ standard deviation.

| | $\bar{\phi}_\omega \pm (\epsilon\%)\bar{\phi}_\omega$ m/s$^2$ | $c_{pw} \pm (\epsilon\%)c_{pw}$ m/s | $A_m \pm (\epsilon\%)A_m$ cm$^2$ | $\nu \pm (\epsilon\%)\nu$ m$^2$/s | $\omega \pm (\epsilon\%)\omega$ 1/s |
|---|---|---|---|---|---|
| $\Delta N_{\omega,\text{sp}}(\epsilon = 5)$ | $0.31 \pm 0.18$ | $0.33 \pm 0.19$ | $0.17 \pm 0.11$ | $0.18 \pm 0.11$ | $0.15 \pm 0.09$ |
| $\Delta N_{\omega,\text{sp}}(\epsilon = 10)$ | $0.61 \pm 0.36$ | $0.66 \pm 0.39$ | $0.35 \pm 0.22$ | $0.35 \pm 0.22$ | $0.30 \pm 0.17$ |
| $\Delta N_{\omega,\text{sp}}(\epsilon = 15)$ | $0.92 \pm 0.54$ | $0.99 \pm 0.59$ | $0.52 \pm 0.32$ | $0.53 \pm 0.33$ | $0.46 \pm 0.25$ |

Fig. S5 shows that the resulting area under the curve (AUC) of the aneurysm physiomarker as a binary predictor for abnormal growth still exceeds 0.99 even for an $\epsilon\% = 5\%$ variation of the input parameters around their measured or assumed constant value (e.g. kinematic viscosity). The AUC drops to 0.98 for an $\epsilon\% = 10\%$ variation, and to 0.94 for $\epsilon\% = 15\%$. Given the maximum change in $\Delta N_{\omega,\text{sp}}$ shown by Table 1 for $\epsilon\% = 5\%$, an uncertainty band of $\pm 0.33$ around the marginal stability case $N_{\omega,\text{sp}} = 0$ can be defined. That is, $N_{\omega,\text{sp}}$ which fall in this band may swing between positive and negative values given natural deviation or measurement error of the physiological input parameters, such as pulse wave velocity $c_p w$. This uncertainty band occupies about 6% of the total range in measured aneurysm physiomarker values $[-2.00, 9.13]$ within the patient cohort. The size of the uncertainty band increases to 12% for $\epsilon = 10$ and 18% for $\epsilon = 15$. Thus, uncertainty in the aneurysm physiomarker scales linearly with measurement error. In scenarios where $N_{\omega,\text{sp}} \approx 0$ falls near the marginal stability state, repeat imaging and more frequent clinical follow-ups are therefore recommended to accurately quantify the physiomarker and predict future abnormal growth for the patient.

5.5.2. *Imaging limitations.* Note that in this study, we have used clinical CT or MR measurements of aortic dimensions to asses growth over time. These measurements are subject to uncertainty due to need for manual selection of an oblique measurement plane, with intra- and inter-observer error rates around 5% in either modality[45;46]. From the prognosis patient cohort, we have conducted a reproducibility analysis on the diameter measurement of SOV and MAA, finding a mean inter-observer error of 4.5% and 4.2% respectively over a set of 35 images each. This agrees with the inter-observer error of 5% reported in literature. Of the 35 SOV images, 5 were from CT; meanwhile, 4 of the 35 MAA images were from CT.

Additionally, there may be considerable discrepancy when different imaging modalities such as MRI or CT are used to measure patients' SOV and MAA diameters at follow-up if measurement standards are not met [47]. However, there appears to be no significant difference in maximum aortic root diameter, ascending aorta diameter, and aortic arch diameter measured using CT and MRI when the same techniques are used (e.g. inner lumen to inner lumen or outer lumen to outer lumen) [48]. Since growth was tracked in this study only at the sinus of valsalva and maximal-area ascending aorta, the low mean differences between imaging modalities (0.2 mm for aortic root, 0.3 mm for the proximal ascending aorta) [48] suggests that growth rates $> 0.2$ cm can be meaningfully distinguished.

To test the effect of measurement error, we introduce 5% Gaussian noise (mean 0%, standard deviation 5%, truncated $\pm 5\%$) to the diameter reported at every time point for all patients. The process was repeated 1000 times, yielding the mean 0.90 and standard deviation 0.02 for the AUC of the aneurysm physiomarker as a binary predictor of abnormal growth $> 0.24$ cm/year. The mean AUC increases to 0.92 when the threshold for abnormal growth is increased to 0.4 cm/year so that the signal threshold exceeds the average noise level. Thus, the performance of the aneurysm physiomarker is robust against the presence of aortic diameter measurement error in this comparative validation.

Note that adding this much random noise can effectively kill any robust measurement of growth through imaging. Even for a very conservative clinical intervention threshold of 0.5 cm/year, a potential intraobserver error of 0.2 cm means that an actual growth rate of 0.1 cm/year can be mistaken for 0.5 cm/year and vice versa. With the input of 5% error, the uncertainty band for direct diameter measurements via imaging is therefore 40% of the range of growth rates; the one order of magnitude lower 6% uncertainty of the aneurysm physiomarker suggests that it outperforms the current clinical standard in tracking abnormal aortic growth.

5.5.3. *Modeling assumptions.* The prediction of flutter in this work is a linear approximation, in the sense that the aneurysm physiomarker measures whether flutter occurs given the patient's current imaged physiological properties. It does not account for changes in physiological properties like blood pressure, aortic stiffness, aortic size, etc. from year to year. In essence, we are positing that observed flutter now is clinically indicative of abnormal growth in the future. This can be ameliorated by more frequent surveillance such as annual or bi-annual imaging for at-risk patients identified via $N_{\omega,\mathrm{sp}} > 0$.

The data analyzed for this study have all originated from one site. Patients were selected without any precondition on inclusion that might explicitly bias the cohort composition in any way with respect to the aneurysm physiomarker; however, future work that investigates predictive performance from prospective data acquired across multiple sites would strengthen confidence for interpretation of our results.

We have conducted a linear stability analysis of a 1D blood vessel model. The immediate advantage is that the problem becomes tractable and yields a closed form solution. However, nonlinear damping or instability inducing effects may become important in certain flow conditions. Note additionally that the flutter instability examined in this work is primarily caused by pressure driven deformation of the aortic wall through the tube law. This flutter is therefore different from the shear induced destabilization of the channel walls induced by the Kelvin-Helmholtz type instability explored in prior literature [18]. Although we have shown that the pressure mediated flutter instability analyzed in this work is strongly associated with abnormal aneurysmal growth, a follow-up study of shear driven wall instabilities [18] is of strong interest, though outside the scope of the current work.

Lastly, the asymmetry, curvature, and branching of the full 3D aortic geometry may also play a role. However, 1D models have in general been well validated against 3D clinical data [49;50;51;30]. The reduced order 1D model we formulate from first principles preserves the key biomechanical features of the actual human aorta – flow pulsatility, wall elasticity, local blood acceleration and shear, etc. – while allowing the system to experience blood-wall interaction instability. The only factor missing is the complex 3D geometry of the system, but it would enter as higher order geometric correction terms to the core biomechanical mechanisms already described. That is, the underlying dimensionless number $N_\omega$ derived would not change because the core physics remain the same.

If the 1D model was not sufficient, then the critical threshold predicted by 1D model would not agree with clinical data. However, good agreement is observed. Experimentally, the 3D character of the actual system is injected in an average sense into our analysis via the actual area and pressure driven acceleration locally measured along the centerline through 4D flow MRI. These local flow values carry the effect of a complex 3D geometry as input into the 1D model. However, anatomical heterogeneity (lumen concavity, convexity, angulation, etc.) could undoubtedly impact the pressure-wall interaction beyond what the flow information can carry into the model. Additional data would be necessary to quantify how less axisymmetric anatomical domains influence generalization of $N_{\omega,\mathrm{sp}}$. This is a challenging task ripe for future work.

## 6. Conclusion

In this work we analyzed an instability-driven growth mechanism of aortic aneurysms from first principles through a linear stability analysis of flow through an elastic blood vessel. The perturbation equations around the base flow gives us a dispersion relation between the temporal growth rate of each flutter mode and its wave number. Floquet theory is used to account for the parametric effect of the heartbeat frequency—essentially, the oscillatory blood flow waveform.

The important parameters determining the onset of unstable flutter—including viscosity, vessel diameter, pressure gradient that drives acceleration, etc.—are collected in a single dimensionless number. Akin to the role of the critical Reynolds number in describing the onset of turbulence, the critical threshold of the dimensionless number tracks the transition of the system to the flutter type instability. If this flutter instability parameter (dimensionless number minus its critical threshold) exceeds zero at a local cross-section of the blood vessel, the growth of perturbation modes may trigger the abnormal dilatation of the local blood vessel. We therefore hypothesize that an aneurysm will form or grow at the site. Otherwise, perturbation amplitudes decay in time, and the location remains stable to this flutter mechanism.

Through follow-up analysis in a group of patients with suspected aortopathy, we've shown that the flutter instability parameter may serve as an aneurysm physiomarker to forecast aneurysm growth. The only input to calculate the aneurysm physiomarker for each patient was a baseline 4D flow magnetic resonance imaging scan taken during the initial visit. We found that this aneurysm physiomarker predicts abnormal aortic growth and/or surgical intervention at clinical follow-up with high accuracy, specificity, and sensitivity.

This ab initio aneurysm physiomarker has the potential to become a predictive diagnostic tool for aneurysm development. It captures the observed qualitative population trends in subjects and clarifies the qualitative growth modes of nascent aortic dilation vs. the evolution of large, developed aneurysms. Here, we have presented a full derivation of the aneurysm physiomarker, tested its potential for diagnostic capability, and contextualized it as a fundamental mechanistic precursor to aneurysm formation and growth.

## 7. Author Contributions

Conceptualization- N.A.P. and T.Y.Z.; planning and supervision- N.A.P., M.M., T.Y.Z., and E.M.I.J; theoretical analysis- N.A.P., T.Y.Z.; clinical methodology- M.M., E.M.I.J., B.D.A., T.Y.Z., S.H., B.C.S. and G.E.; writing- N.A.P., M.M., T.Y.Z., G.E., E.M.I.J., and S.H.

## 8. Competing interests

The authors have no competing financial interests or other interests that might be perceived to influence the results and/or discussion reported in this paper.

## 9. Acknowledgements

Research reported in this publication was supported by the National Heart, Lung, And Blood Institute of the National Institutes of Health under Award Number F32HL162417. The content is solely the responsibility of the authors and does not necessarily represent the official views of the National Institutes of Health.

| | normal subjects ($n = 100$) | all aortopathy patients ($n = 117$) | p-value | prognosis aortopathy patients ($n = 72$) |
|---|---|---|---|---|
| age (years) | 46.2±15.5 [19,79] | 57.4±14.2 [22,86] | $2 \times 10^{-7}$ | 58.6±11.9 [29,79] |
| sex (female) | 50 [50%] | 35 [30%] | $1.5 \times 10^{-3}$ | 17 [24%] |
| height (m) | 1.71±0.11 [1.30,1.96] | 1.73±0.15 [1.14,2.03] | $7.4 \times 10^{-3}$ | 1.76±0.13 [1.40,2.03] |
| weight (kg) | 79.2±17.9 [47.6,142.9] | 84.8±18.1 [45.5,140.9] | $3.5 \times 10^{-2}$ | 86.5±19.2 [45.5,140.9] |
| follow-up (years) | - | - | - | 5.86 ± 1.77 [1.13,8.67] |

TABLE 3. Characteristics of the study cohort are summarized as mean ± standard deviation and [minimum,maximum] of range or [percentage] values. Prognosis aortopathy patients had clinical follow-up data and were analyzed for growth rate validation of the aneurysm physiomarker; other patients did not have follow-up data. The p-values are reported for a Wilcoxon rank sum test between cohort statistics of healthy subjects and all aortopathy patients.

TABLE 4. Each physiological term that contributes to measuring the aneurysm physiomarker $N_{\omega,\mathrm{sp}}$ is tabulated for both patients and normal subjects in three age groups. This includes the patient's aorta diameter $A_m$, blood pressure gradient $\bar{\phi}_\omega$ causing oscillatory accelerations, pulsatile contribution to wall shear $\beta_b$, heartbeat angular frequency $\omega$, and pulse wave velocity $c_{pw}$. The one-tailed Wilcoxon rank sum test was used to determine whether the larger median of one population (e.g. patients, $N_{\omega,\mathrm{sp}} > 0$) is significantly greater than the smaller median of the other (e.g. patients, $N_{\omega,\mathrm{sp}} \leq 0$). The row of p-values comparing patient and normal subject cohorts is colored red, while the row of p-values comparing $N_{\omega,\mathrm{sp}} > 0$ and $N_{\omega,\mathrm{sp}} \leq 0$ is colored blue. Rejection of the null hypothesis at the 5% level is colored orange.

| | $N_{\omega,\mathrm{sp}} > 0$ | | | | | $N_{\omega,\mathrm{sp}} \leq 0$ | | | | |
|---|---|---|---|---|---|---|---|---|---|---|
| | $\bar{\phi}_\omega$ m/s$^2$ | $c_{pw}$ m/s | $A_m$ cm$^2$ | $\beta_b$ none | $\omega$ 1/s | $\bar{\phi}_\omega$ m/s$^2$ | $c_{pw}$ m/s | $A_m$ cm$^2$ | $\beta_b$ none | $\omega$ 1/s |
| **Patients (Age$<$40)** | | | | | | | | | | |
| median | 7.7377 | 5.2650 | 7.5719 | 42.0645 | 7.1400 | 7.7079 | 7.3322 | 5.8507 | 38.2583 | 8.3087 |
| p-value between $N_{\omega,\mathrm{sp}}$ | 0.4699 | 0.0014 | 0.0168 | 0.13006 | 0.3706 | - | - | - | - | - |
| p-value between cohorts | 0.2648 | 0.4666 | 0.0358 | 0.1572 | 0.3144 | 0.0113 | 0.0113 | 0.3517 | 0.5218 | 0.4008 |
| **Normal subjects** | | | | | | | | | | |
| median | 7.5684 | 5.0998 | 6.4396 | 39.7752 | 8.0143 | 6.2113 | 6.3452 | 5.6152 | 36.9418 | 8.4361 |
| p-value between $N_{\omega,\mathrm{sp}}$ | 0.0081 | $5 \times 10^{-6}$ | 0.0533 | 0.1054 | 0.4539 | - | - | - | - | - |
| **Patients ($40 \leq$ Age $< 60$)** | | | | | | | | | | |
| median | 7.1694 | 5.3235 | 9.0482 | 43.0926 | 8.1812 | 6.4033 | 7.8868 | 8.9853 | 44.5748 | 7.7121 |
| p-value between $N_{\omega,\mathrm{sp}}$ | 0.1230 | $3 \times 10^{-8}$ | 0.3522 | 0.17438 | 0.4378 | - | - | - | - | - |
| p-value between cohorts | 0.0966 | 0.0335 | 0.0526 | 0.1905 | 0.2701 | 0.2821 | 0.0528 | 0.0153 | 0.0018 | 0.5028 |
| **Normal subjects** | | | | | | | | | | |
| median | 9.1976 | 6.0647 | 7.1551 | 40.2202 | 8.3585 | 7.2424 | 7.5973 | 7.7359 | 40.4069 | 7.8228 |
| p-value between $N_{\omega,\mathrm{sp}}$ | 0.0092 | 0.0286 | 0.3518 | 0.4939 | 0.2515 | - | - | - | - | - |
| **Patients (Age $\geq 60$)** | | | | | | | | | | |
| median | 6.6368 | 4.2547 | 10.2915 | 43.5787 | 7.5166 | 6.7577 | 8.5646 | 10.1875 | 43.5171 | 7.4592 |
| p-value between $N_{\omega,\mathrm{sp}}$ | 0.3181 | $6 \times 10^{-9}$ | 0.2232 | 0.3592 | 0.4268 | - | - | - | - | - |
| p-value between cohorts | - | - | - | - | - | 0.3190 | 0.2551 | 0.0039 | 0.0107 | 0.1911 |
| **Normal subjects** | | | | | | | | | | |
| median | - | - | - | - | - | 6.9007 | 8.3760 | 8.5243 | 39.8121 | 7.7121 |
| p-value between $N_{\omega,\mathrm{sp}}$ | - | - | - | - | - | - | - | - | - | - |

## S1. Supplementary

S1.1. **Governing equations - relationship between Womersley number and wall shear.** For a parabolic velocity profile corresponding to Womersley number $w_o = 0$, the wall shear coefficient $\beta = 8$. As $w_o$ increases, the shear contribution becomes localized in a boundary layer at the wall, leading to a larger value of $\beta$ for higher frequency flow through the blood vessel. Figure S1 presents $\beta$ normalized by a factor of 8 as $w_o$ varies. The initial nonlinear behavior for $w_o < 2$ smooths out into a linear relation when $w_o > 2$ and the transient inertial forces are large. This relationship spans the full range of physiological heartbeat frequencies.

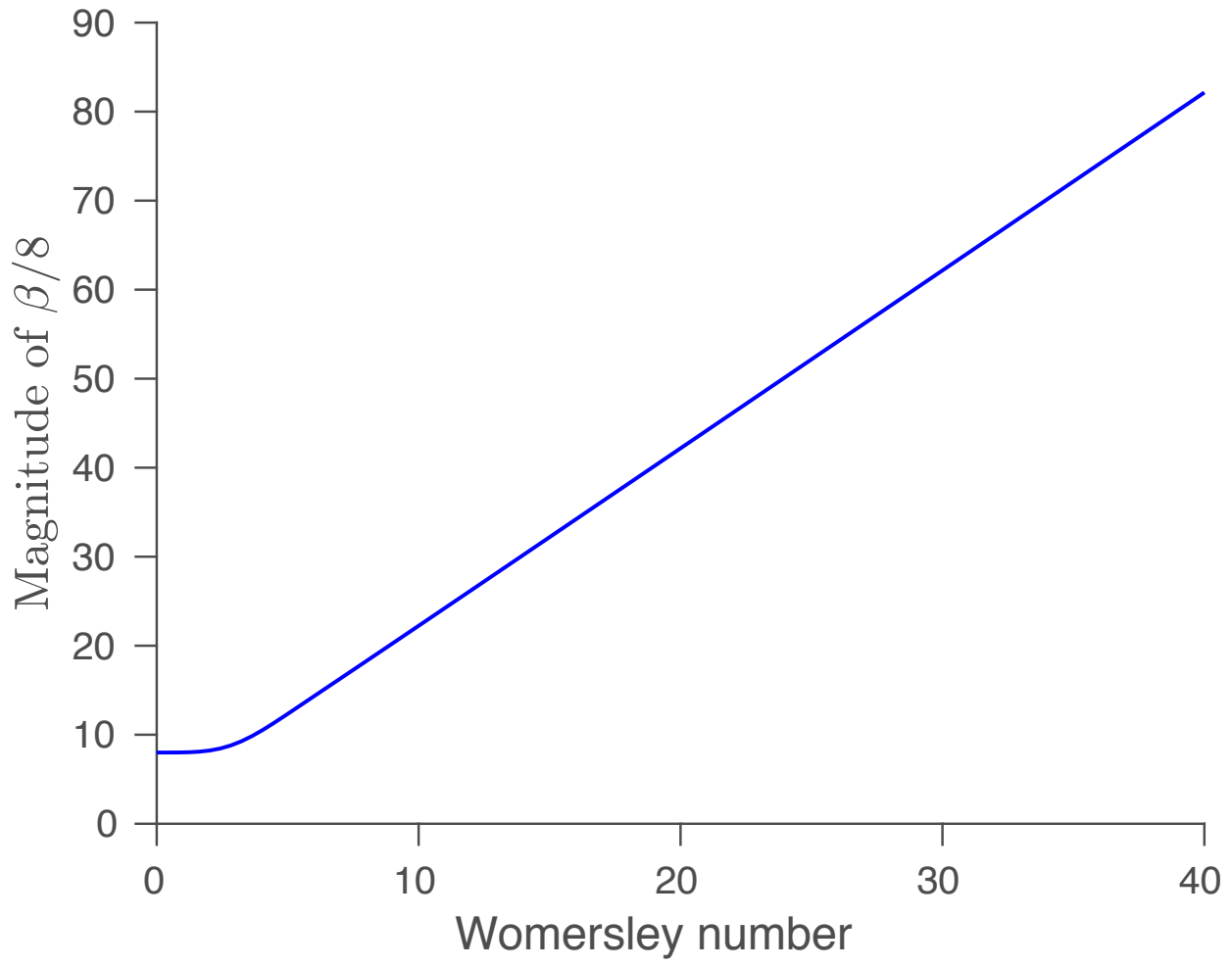

FIGURE S1. The viscous factor $\beta$ as a function of Womersley number $w_0$. Here, $\beta$ has been normalized by its value at $w_0 = 0$, which corresponds to a parabolic velocity profile.

S1.2. **Pulse wave velocity - relationship with aortic wall stiffness.** The relationship between aortic wall stiffness and pulse wave velocity can be derived by transforming the set of simplified governing equations to the standard form of the wave equation[1]. The relevant conservation equations are

$$\frac{\partial A}{\partial t} + \frac{\partial Au}{\partial x} = 0, \tag{S1}$$

$$\frac{\partial u}{\partial t} + u\frac{\partial u}{\partial x} = -\frac{\partial P}{\partial x}, \tag{S2}$$

where the viscous term has been neglected, and $P$ is the dynamic pressure divided by the blood density. A general tube law is used

$$P = \frac{1}{\rho}G(A), \tag{S3}$$

where $G$ is some function of the local cross-sectional area. The function $G$ represents the full dependence of the excess internal pressure to the cross-sectional area and thus can encapsulate aortic wall properties such as elastic moduli, wall thickness, etc. in the most general case. Without adding new notation, we next introduce an invertible change in the independent variables $x \to x + vt$ and $t \to t$ where the the velocity $v$ is frozen at the mean value. In the new basis, the conservation equations become

$$\frac{\partial A}{\partial t} + A\frac{\partial u}{\partial x} = 0, \tag{S4}$$

$$\frac{\partial u}{\partial t} + \frac{1}{\rho}\frac{\mathrm{d}G}{\mathrm{d}A}\frac{\partial A}{\partial x} = 0. \tag{S5}$$

Differentiating the mass equation (eqn. S4) with respect to time and the momentum equation (S5) with respect to space gives

$$\frac{\partial^2 A}{\partial t^2} + A\frac{\partial^2 u}{\partial x \partial t} = 0, \tag{S6}$$

$$\frac{\partial^2 u}{\partial x \partial t} + \frac{1}{\rho}\frac{\mathrm{d}G}{\mathrm{d}A}\frac{\partial^2 A}{\partial x^2} = 0, \tag{S7}$$

which can be combined to obtain

$$\frac{\partial^2 A}{\partial t^2} - \Big(\frac{1}{\rho}\frac{\mathrm{d}G}{\mathrm{d}A}A\Big)\frac{\partial^2 A}{\partial x^2} = 0. \tag{S8}$$

This is the standard form of the wave equation, where the term in parenthesis is typically called the propagation speed. It represents the speed of the plane wave solutions to eqn. S8. The pulse wave velocity can thus be defined as

$$c_{pw}^2 = \frac{1}{\rho}\frac{\mathrm{d}G}{\mathrm{d}A}A \tag{S9}$$

S1.3. **Dimensionless groups.** Dimensionless groups are introduced in Table S1. The resulting dimensionless number $N_\omega$ captures the dominant physical drivers and inhibitors of the flutter instability.

Table S1. From the basic variable inputs in the first row, the natural length $L$ and time $T$ scales of the system in the middle row are introduced to produce the dimensionless groups in the last row.

| | | | | | | | | |
|---|---|---|---|---|---|---|---|---|
| **Basic input variables** | | | | | | | | |
| $\nu$ $\mathrm{m^2/s}$ | $\omega$ $\mathrm{1/s}$ | $R_m$ $\mathrm{m}$ | $K_e$ $\mathrm{kg/(m \cdot s^2)}$ | $A_o$ $\mathrm{m^2}$ | $u_m$ $\mathrm{m/s}$ | $u_\omega$ $\mathrm{m/s}$ | $\phi_m$ $\mathrm{m/s^2}$ | $\bar{\phi}_\omega$ $\mathrm{m/s^2}$ |
| **Natural length and time scales of the system** | | | | | | | | |
| | | | $L = \sqrt{A_m}$ | | $T = \frac{2A_m}{\beta_b \pi \nu}$ | | | |
| **Dimensionless groups** | | | | | | | | |
| $\tilde{\mu} = T\mu$ | $\tilde{\omega} = T\omega$ | $N_T = \frac{K_e {A_m}^2}{\rho A_o (\frac{\beta_b}{2}\pi\nu)^2}$ | $k'' = kL\sqrt{N_T}$ | $\tilde{A} = \frac{\hat{A}}{L^2}$ | $u'' = \frac{\hat{u}T}{L\sqrt{N_T}}$ | $N_m = \frac{\phi_m T^2}{L\sqrt{N_T}}$ | $N_\omega = \frac{\bar{\phi}_\omega T^2}{L\sqrt{N_T}}$ | |

S1.4. **Cohort criteria.** Here, we present our thoughts on the cohort selection and inclusion / exclusion criteria.

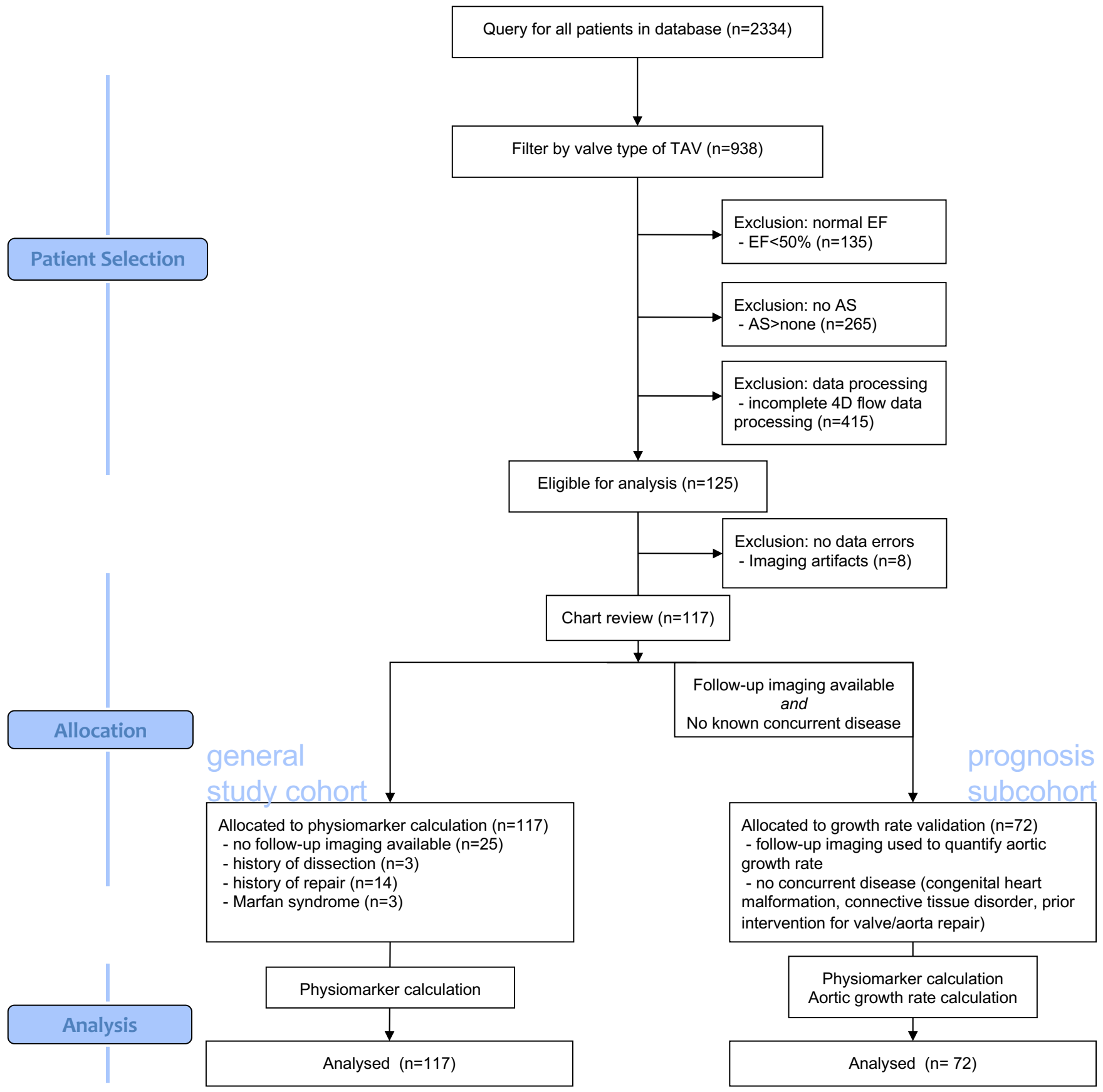

FIGURE S2. A flow diagram for inclusion / exclusion. The diagram reflects a two-stage process for selecting patients, which resulted in the two different analysis groups described in the manuscript. The database was queried to identify patients matching the general inclusion/exclusion criteria (TAV, EV$\geq$50%) with data available for analysis (phase corrections applied to images, aortic segmentation created), thus creating the general cohort of patients. A chart review was conducted to identify patients with no genetic tissue disorders or history of surgical intervention as well as to tabulate aortic dimensions and potential later surgical intervention on follow-up.

S1.4.1. *Patient cohort.* The patient cohort was drawn from a Radiology Department archive of imaging data and metadata that captures scans from patients who were referred for clinical 4D flow MR imaging. Each record includes the 4D flow MRI of the thoracic aorta, in addition to basic clinical status data (valve type, AS/AR, SOV/MAA diam., EF) and demographic data (age, sex). For some records, the database also includes data files containing the velocity fields and aortic segmentations derived from processing the raw 4D flow images (not all subject data have been processed).

A diverse group of patients was included in the initial cohort formation to allow evaluation of the aneurysm physiomarker across a wider range of patients. The flutter instability parameter's relationship with growth is not specific to a disease type or state; it is derived solely from physical analysis of the blood-wall interaction instability. Therefore, it is not a confound for this study to include patients with different etiology of aortopathy when evaluating general differences relative to a healthy population.

Note that the initial exclusion criteria were structured to avoid known factors for aortic growth (such as aortic stenosis AS or heart failure HF), which could have dominating influence on growth trends by being strong drivers of hemodynamic derangement. Instead, we chose to tackle the more difficult problem by sampling a patient cohort that precluded these overt indicators of growth. Moreover, for exclusion of ejection fraction EF $< 50\%$, only a small number of patients were excluded as a result (less than 15% of all TAV patients, see Fig. S2).

However, in the follow-up subcohort for evaluating prognostic potential for the aneurysm physiomarker, we focused on patients with no genetic tissue disease, etc., as such patients receive a different clinical management regimen than do dilatation patients without such disorders.

S1.4.2. *Healthy cohort.* The healthy volunteers were recruited as part of a separate study to obtain normative measurements for hemodynamic parameters in healthy subjects. A total of 242 healthy subjects had been recruited with data available at the time of this study, from which 100 were selected so as to include 10 males and 10 females with no image artifacts in each age grouping of 19-30, 31-40, 41-50, 51-60, and 61-79 years. The selection of 100 subjects for analysis was performed by including the first 10 recruited of each sex in each age grouping. For example, if a 38 year-old female was recruited in 2018, but ten females between 31 and 40 years old had already been recruited in 2011-2017, the 2018-recruited 38 year-old female would not be included, while the ten previously-recruited females would be.

Note from Table 3 that the 'healthy' vs 'all aortopathy' cohorts were not age/sex/weight/height matched. As future work, a thorough comparative study with cohort matching should be carried out to validate the proposed aneurysm physiomarker, drawing upon multi-center imaging and follow-up data. The main results of this study are to show that the aneurysm physiomarker is predictive (via follow-up analysis) and that it captures potential growth instabilities that can drive aneurysmal development in healthy subjects. In the context of this study, whether the difference in abnormal growth experienced by patient vs healthy subject cohorts arise from differences in age/sex/weight/height is not the focus, since prior work has been plentiful in showing the trends between aortic aneurysm growth with age/sex/weight/height through statistical correlations.

S1.5. **Image Acquisition and Preprocessing.** Clinical imaging was performed at 1.5T and 3T (Aera / Avanto / Espree, Siemens, Germany). Sequence parameters for 4D flow MRI included 1.2–3.1 $\times$ 1.2–3.1 $\times$ 1.2–5.0 mm$^3$ / 33–45 ms spatial / temporal resolution; 12.4–40.6 $\times$ 18.0–50.0 $\times$ 3.8–17.6 mm$^3$ field of view; 80–500 cm/s VENC, as appropriate, determined from flow scout image; 2.1–3.0 ms TE, 4.1–5.7 ms TR, 7–25° tip angle; and respiratory navigators for free-breathing scans. Scans for patients used a contrast agent (Ablavar, Magnevist, Multihance), and scans for healthy subjects did not. Images for all subjects were acquired between January 2011 and December 2019. Pre-processing of 4D flow MRI data included previously-described methods for correction of background phase from eddy currents and Maxwell terms and for velocity phase un-aliasing [2]. Preprocessing was performed with commercial computational software (MATLAB, Mathworks, Natick, Massachusetts). Following preprocessing, a three-dimensional phase-contrast MR angiography (3D PCMRA) was generated as a starting point for aortic segmentation by calculating time-averaged velocity sum-of-squares. The 3D PCMRA was then opened in commercial image processing software (Mimics Innovation Suite, Materialize, Leuven, Belgium) along with magnitude images and edited to remove errant inclusions or add missed sections of the aorta.

Note that we did not find any significant difference in acquisition parameters such as magnetic field strength, flip angle, pixel bandwidth, echo time, repetition time, slice thickness, VENC, spatial resolution, temporal resolution, etc. ($p>0.2$) between patients with abnormal vs low growth as defined by the aneurysm physiomarker $N_{\omega,\text{sp}} > 0$ and vice versa. This statistical comparison was done with the Wilcoxon rank sum test. Likewise, there was also no significant difference between high vs low growth patients in use or non-use of contrast agents (p=0.9).

S1.6. **Image Processing.** The 4D flow MRI provides information about the three-dimensional geometry of the aorta as well as the velocity field inside it. The 3D geometry is time-averaged from velocity contrast, and therefore remains constant in time. Fig. S3(a) shows the time-averaged geometry of an aorta from a 4D flow MRI. The geometry is generated on a Cartesian grid of voxels which have a binary value, i.e., voxels lying outside and inside the aorta have a value of 0 and 1, respectively. The measured velocity field is a function of both space and time.

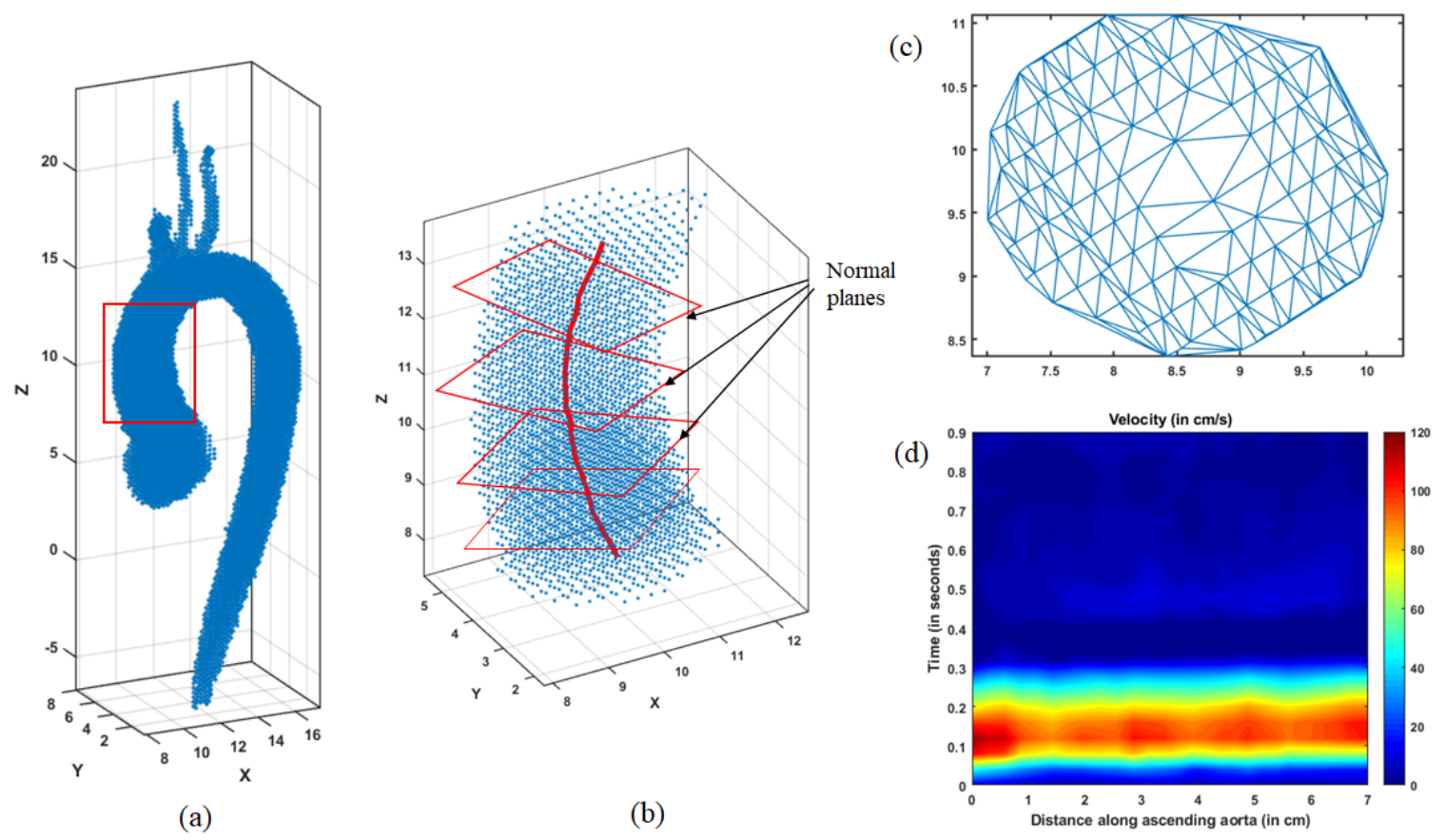

FIGURE S3. Cross-sectional areas and mean velocity field from 4D flow MRI of the aorta. (a) Time-averaged 3D geometry of the aorta. The red box marks the ascending aorta. The axes units are in cm; (b) Point cloud (in blue) showing the ascending aorta. The red curve shows the centerline, and the red boxes show the planes normal to the centerline. These planes are used to calculate the cross-sectional areas and mean velocities. The axes units are in cm; (c) An example of the aorta cross-section on a normal plane. Meshing is done using Delaunay triangulation to calculate the cross-sectional area at the normal plane. The axes units are in cm; (d) Variation of mean velocity as a function of time and length along the ascending aorta.

Our analysis was focused only on the ascending aorta, from the aortic root to just below the three branches at the aortic arch. This region is shown inside the red box in Fig. S3 (a) and more clearly zoomed in Fig. S3(b). The upper and lower limits of the ascending aorta were segmented manually.

To model the ascending part of the aorta in a one-dimensional model, we find the variation of cross-sectional area and mean velocity along its length. A centerline is first generated through the ascending aorta (Fig. S3(b)). Normal planes are then generated. These planes were used to calculate the cross-sectional areas and mean velocities at every point along the centerline. Voxels on each plane are then meshed using Delaunay triangulation (Fig. S3(c)). The sum of these triangles is the cross-sectional area $A_m$ of the aorta at a particular centerpoint.

The mean velocity at each cross-section are calculated by

$$u(x,t) = \frac{1}{N}\sum_{i}^{N} \mathbf{v}_i(x,t)\cdot\hat{\mathbf{n}}, \tag{S10}$$

where $u$ is the mean velocity at the centerline, $i$ represents the $i$-th point in plane, $N$ is the total number of points, $\mathbf{v}_i$ is the velocity at the $i$-th point, $\hat{\mathbf{n}}$ is the unit normal to the plane, $x$ is the distance along the centerline, and $t$ is time. The variation of $u(x,t)$ during a cardiac cycle is shown in Fig. S3(d).

These velocity profiles can be used to calculate the pressure gradient that drives oscillatory blood acceleration $\bar{\phi}_\omega$ via eqn. 10 as well as the pressure gradient that drives the mean blood acceleration $\phi_m$ via $u_m = \frac{\phi_m A_m}{\beta_m \pi \nu}$. The pulse wave velocity $c_{pw}$ is also found from the velocity profiles $u(x,t)$ using the cross correlation (XCor)[3] method, which agrees with literature values for healthy controls and patients[5]. Briefly, the XCor PWV value is calculated by automated placement of a centerline in the aortic segmentation and creation of analysis planes every 4mm along the length, then cross-correlating the through-plane flow-time curves to find the transit time between all locations of the aorta[6;7]. The heartbeat frequency $\omega$ is used to calculate the Womersley number $w_0 = R\sqrt{\omega/\nu} \geq 0$ and then the friction coefficient $\beta_b$ from eqn. 9. The remaining parameters of kinematic viscosity $\nu$ and density $\rho$ of blood used to determine $N_{\omega,\mathrm{sp}}$ were sourced from reference values in literature ($\nu \approx 4e - 3\mathrm{N\ s/m^2}$ and $\rho \approx 1060\ \mathrm{kg/cm^3}$)[8]

S1.7. **Age & sex dependencies of the aneurysm physiomarker.** Table S2 shows that the distribution of the aneurysm physiomarker among different sex and age groups in the two cohorts appear to agree with general population trends reported in the literature.

| | $N_{\omega,\text{sp}}$ | | |
|---|---|---|---|
| | Age $<$ 40 | 40 $\leq$ Age $<$ 60 | Age $\geq$ 60 |
| **Patients** | | | |
| median (female) | 2.8590 | 0.3372 | -0.1557 |
| p-value between sexes | 0.0531 | 0.1837 | 0.3264 |
| p-value (female) between age groups | 0.0388 | 0.1764 | 0.0091 |
| p-value (female) between cohorts | 0.0042 | 0.0952 | 0.0227 |
| median (male) | 0.3330 | -0.1027 | -0.0963 |
| p-value (male) between age groups | 0.1002 | 0.4582 | 0.1043 |
| p-value (male) between cohorts | 0.4571 | 0.2944 | 0.0150 |
| **Normal subjects** | | | |
| median (female) | -0.1244 | -0.0660 | -0.6831 |
| p-value between sexes | 0.0122 | 0.3276 | 0.4849 |
| p-value (female) between age groups | 0.4037 | $6 \times 10^{-4}$ | 0.0018 |
| median (male) | 0.7286 | -0.1000 | -0.8908 |
| p-value (male) between age groups | 0.0019 | 0.0080 | $3 \times 10^{-4}$ |

TABLE S2. The aneurysm physiomarker $N_{\omega,\text{sp}}$ stratified by age and sex. The one-tailed Wilcoxon rank sum test was used to determine whether the larger median of one population (e.g. patients, Age $<$ 40, female) is significantly greater than the smaller median of the other (e.g. normal subjects, Age $<$ 40, female). The row of p-values comparing patient and normal subject cohorts is colored red, while the row of p-values comparing sexes is colored blue. The row of p-values comparing each age group is colored green; note that the p-value beneath Age $<$ 40 tests the age groups Age $<$ 40 and 40 $\leq$ Age $<$ 60, the p-value beneath 40 $\leq$ Age $<$ 60 tests the age groups 40 $\leq$ Age $<$ 60 and 40 $\leq$ Age, and the p-value beneath Age $\geq$ 60 tests the age groups Age $\geq$ 60 and Age $<$ 40. Rejection of the null hypothesis at the 5% level is colored orange.

S1.8. **The changes in aortic wall compliance during aneurysm development.** Changes in aortic wall stiffness has been shown in the literature to determine the trajectory of aneurysm progression, as shown in Table S3. Specifically, wall stiffening appears to result in stable aneurysms that do not exhibit increased risk of rupture or significant growth (Type 1), while compliant walls are associated with unstable aneurysms that are at increased risk of further abnormal dilatation and rupture (Type 2).

Table S3. Summary of relevant published literature on the role of aortic wall distensibility in aneurysm progression and rupture.

| Aneurysm studied | Method/modality | Summary of findings | Author/year |
|---|---|---|---|
| AAA | 112 patients with initially non-operated AAA were recruited from five centres. They underwent baseline compliance measurements and were then followed for a median of 7 months. | • Comparing AAA of similar sizes, AAA which rupture or require elective repair appear to be more compliant than those AAA that do not.<br>• Aneurysms can be classified into two types (**a**) Type I - further enlargement is accompanied by increasing stiffness from increased collagen deposition and/or remodelling in the aortic wall. This confers strength to the AAA so that the risk of rupture is low.<br>• (**b**) Type II - further enlargement is not associated with an increasing stiffness, and stiffness may even fall. This can result from a failure to lay down and remodel collagen, leading to weak or "thinning" aortic walls. These aneurysms are at risk of rupture. | Wilson et. al (1998) [9] |
| AAA | 62 males of median age 68 with detected AAA were screened for circulating markers of elastin and collagen metabolism | • Increased elastolysis- which induces media degradation and leads to aneurysm rupture- is associated with increasing distensibility of the aortic wall<br>• Most aneurysms become less distensible as they expand; those that fail to grow stiffer or become suddenly more distensible are at high risk of rupture.<br>• The change in distensibility within each aneurysm is of greater significance than differences between a level of “normal” distensibility and that of the aneurysm. | Wilson et. al (2001) [10] |
| AAA | A prospective, six-center study of 210 patients with AAA was conducted. Blood pressure (BP), maximum AAA diameter (Dmax), and AAA distensibility (pressure strain elastic modulus Ep) were measured at 6 months with an ultrasound scan–based echo-tracking technique. | • The reduction in AAA distensibility over time is associated with a significantly shorter time to rupture, independent of other risk factors (age, sex, BP, Dmax). | Wilson et. al (2003) [11] |
| AAA | A prospective study of 56 patients with AAA was conducted using tissue Doppler imaging system. | • There is a significant positive relationship between maximum diameter and the segmental compliance of the aneurysm.<br>• When stratified by size into two groups (group 1, AAA diameter $<$ 45 mm and group 2, AAA diameter $\geq$ 45 mm), group 2 had significantly higher segmental compliance while group 1 exhibited greater scatter in stiffness. | Long et. al (2005) [12] |
| AAA | A study of 43 patients with infrarenal AAAs was conducted in the postoperative period | • Patients with electively repaired AAAs have accelerated pulse wave velocities, indicating highly rigid aortic walls.<br>• Patients with ruptured aneurysms exhibited significantly lower pulse wave velocities with greater variance and scatter.<br>• Patients with high aortic compliance experience faster growth and earlier rupture. | Russo (2006) [13] |
| TAA | A study of 32 patients with ascending TAAs and 46 age matched controls was conducted to measure the femoral pulse wave velocity (cfPWV), heart-femoral pulse wave velocity (hfPWV) and brachial-ankle pulse wave velocity (baPWV) | • In patients with ascending TAAs, there was a significant inverse relationship between aortic diameter at the SOV and cfPWV, as well as hfPWV, but not with baPWV. This correlation was not present in controls without ascending TAA. | Rabkin et. al (2014) [14] |
| TAA | A study of 40 patients with TAAs was conducted to measure regional aortic diameter and PWV using 1.5 T MRI | • Incidence of increased regional PWV exhibited moderate specificity but low sensitivity for coexisting with regional aortic dilatation in the ascending aorta and aortic arch. | Kröner et. al (2015)[15] |
| AAA | Biaxial extension tests, second-harmonic generation imaging and histology were performed on 15 samples from the anterior part of AAA walls harvested during open aneurysm surgery. | • Three stages of disease progression were identified. (**a**) Stage 1 - intimal thickening is accompanied by a decrease in elastin and smooth muscle cell (SMC) content. Stiffness decreased by a factor of 2 compared to a healthy aorta.<br>• (**b**) Stage 2 - Further decrease in elastin and SMC content along with increase in adipocytes in the wall. A neo-adventitia layer formed from new collagen deposition on the outer AAA walls, which did not appear in healthy or stage 1 subjects. Wall distensibility increases compared to healthy or stage 1 AAA subjects. .<br>• (**b**) Stage 3 - Significant buildup of neo-adventitia occurs along with media and intima breakdown. Two types of stage 3 walls were observed. Type 1 AAA had 'safely' remodeled walls with no adipocytes present in the wall and a thick collagen layer. Type 2 AAA remodeled to a 'vulnerable' state, exhibiting significant inflammation and adipocytes inside the wall. | Niestrawska et. al (2019)[16] |

S1.9. **Aneurysm physiomarker trend with patient increased growth rates.** Fig. S4 demonstrates a positively proportional relationship between the increasing aneurysm physiomarker $N_\omega$ and greater growth rates in the SOV and MAA. The correlation coefficient between $\Delta$ $\text{SOV}_{\max}$ and the aneurysm physiomarker is 0.56, with a p-value of $4 \times 10^{-7}$, whereas the correlation coefficient between $\Delta$ $\text{MAA}_{\max}$ and the aneurysm physiomarker is 0.58, with a p-value of $8 \times 10^{-8}$.

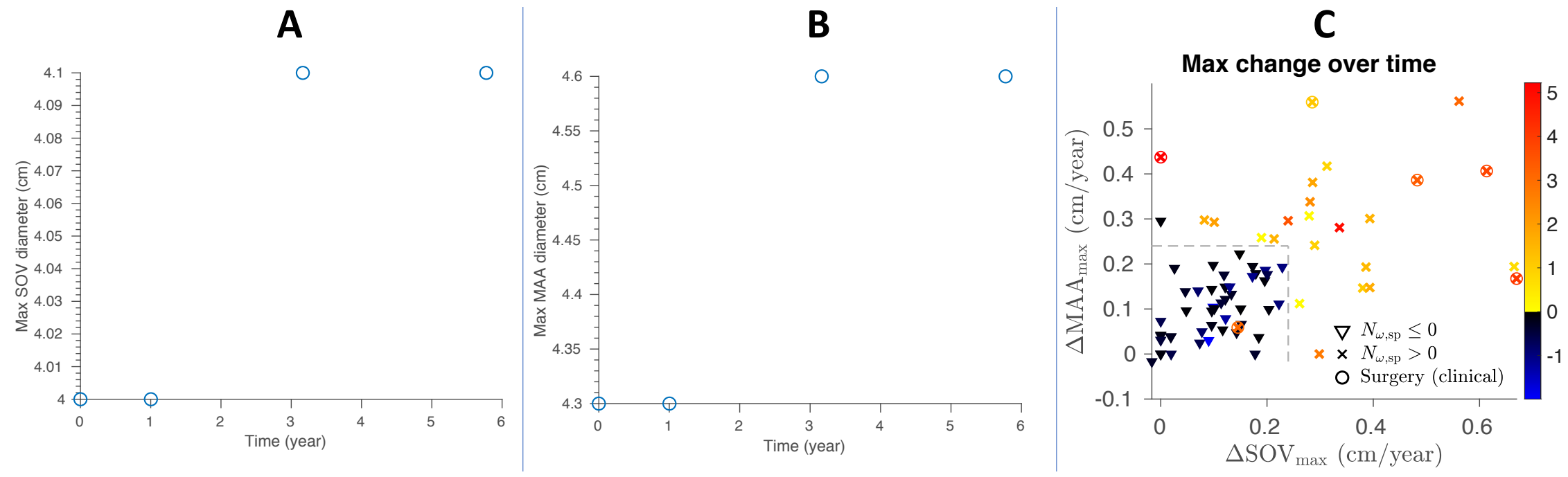

FIGURE S4. **A**) Example of the maximum SOV diameter recorded during each clinical visit for one patient. **B**) Example of the maximum MAA diameter recorded during each clinical visit for one patient. **C**) A prediction vs outcome diagram of all patients with follow-up imaging data. The maximum growth rate of their MAA and SOV in (cm/year) are simultaneously visualized by color with respect to the magnitude of $N_{\omega,\text{sp}}$. $N_{\omega,\text{sp}}$ is calculated from an MRI at time zero. If $N_{\omega,\text{sp}} > 0$, the patient's marker is labeled by x's. Otherwise, the data point is labeled by downward pointing triangles. The circles indicate that the patient experienced a surgical intervention after their initial MRI at year 0. $N_{\omega,\text{sp}} > 0$ appears to correlate with larger growth rates for the MAA and SOV. The growth threshold of 0.24 cm/year is labeled by black dotted lines- this value is outside the normal range of growth in TAAs of all sizes[17] (< 0.2 cm/year) and optimally discriminates between stable and unstable aneurysms predicted by the proposed aneurysm physiomarker. This optimal threshold of 0.24 cm/year falls within the clinically observed range of abnormal growth (0.24 cm/year for small aneurysms to 0.31 cm/year or large aneurysms) that is associated with chronic dissection.

S1.10. **Additional limitations.** We estimate the sensitivity of the aneurysm physiomarker performance to uncertainty in the input physiological terms. By using the aortic growth rate of 0.24 cm/year as an indicator of significant growth, the aneurysm physiomarker $N_{\omega,\mathrm{sp}} > 0$ serves as a good binary predictor for the growth outcome of each patient. The area under the curve (AUC) of a receiver operating characteristic (ROC) analysis is 0.997 for the unaltered input parameters and 0.994 for a 5% variation of the input parameters around their measured or assumed constant value (e.g. kinematic viscosity of blood).

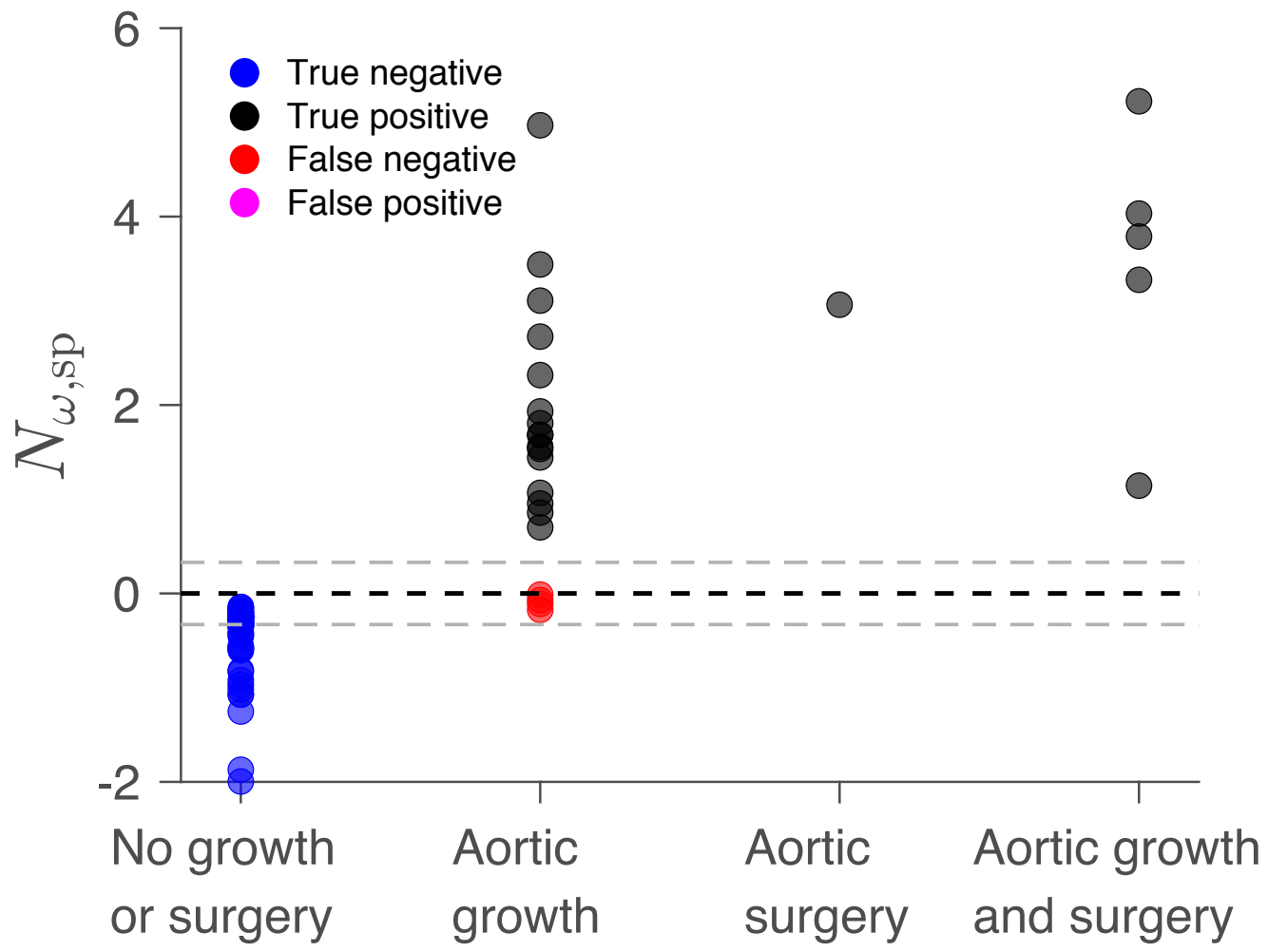

FIGURE S5. After varying each input parameter independently by 5% around their measured or assumed constant value (kinetmatic viscosity), the largest resulting change in the aneurysm physiomarker which pushes $N_{\omega,\mathrm{sp}}$ toward the opposite sign is plotted. Each patient has been labeled according to whether $N_{\omega,\mathrm{sp}} > 0$ accurately predicts a growth outcome, categorized as exhibiting a growth rate in SOV or MAA $\geq 0.24$ cm/year or experiencing surgical intervention at follow-up. A 5% change in the input parameters can induce at maximum a difference of 0.33 in the aneurysm physiomarker $N_{\omega,\mathrm{sp}}$; this uncertainty band has been labeled by the grey dotted lines around the marginal stability threshold of $N_{\omega,\mathrm{sp}} = 0$. That is, $N_{\omega,\mathrm{sp}}$ which fall in this band of $\pm 0.33$ are sensitive (may swing between positive and negative values) to errors in the measurement of the physiological input parameters, such as pulse wave velocity. Repeat imaging and more frequent clinical follow-ups are therefore recommended to accurately quantify the physiomarker and predict future abnormal growth.

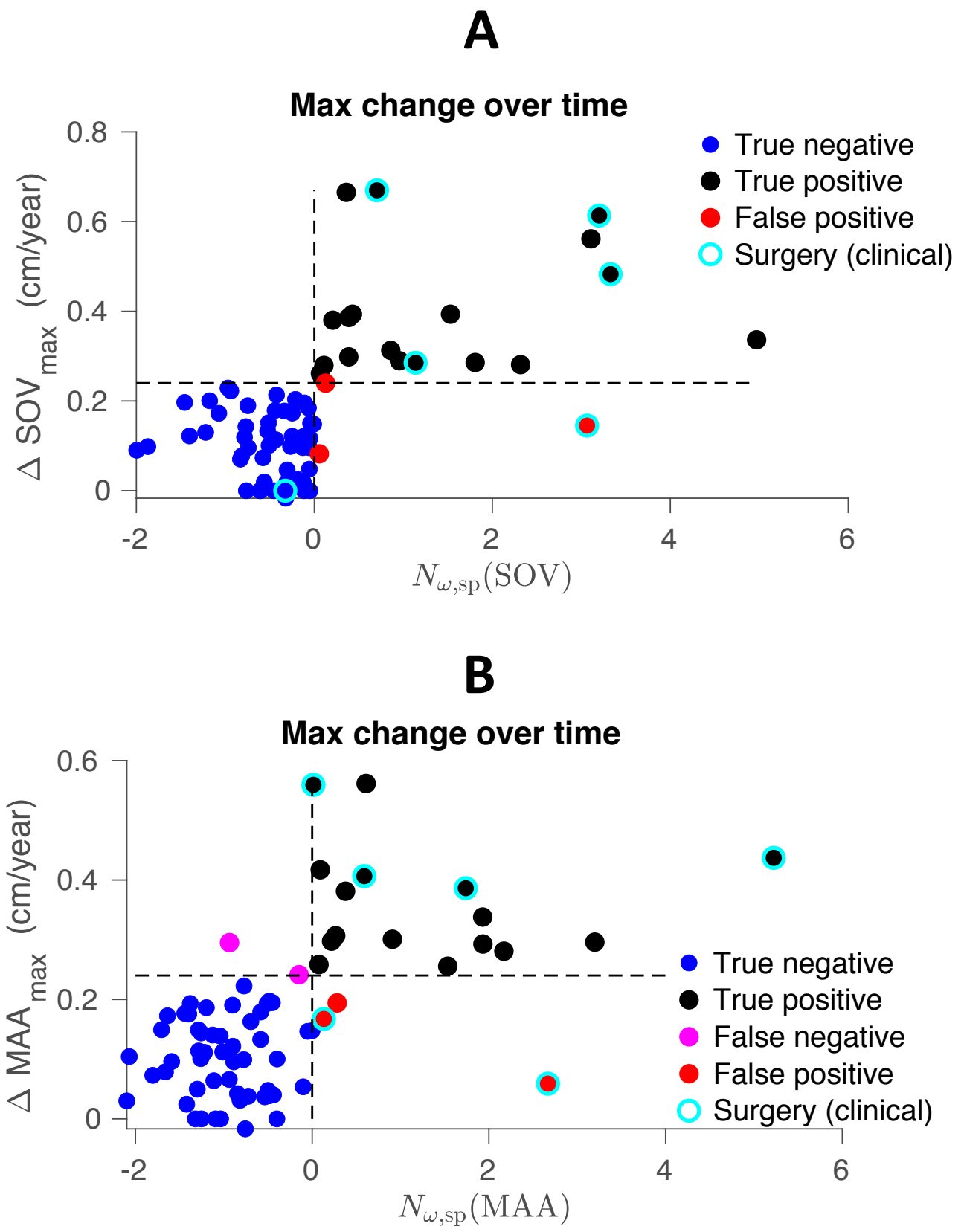

Figure S6. **A**) The growth rate at the SOV vs the local aneurysm physiomarker, defined as the maximum $N_{\omega,\mathrm{sp}}$ along the beginning 20% of the ascending thoracic aorta (truncated before the arch). **B**) The growth rate at the MAA vs the local aneurysm physiomarker, defined as the maximum $N_{\omega,\mathrm{sp}}$ along the middle 40% to 60% of the ascending thoracic aorta (truncated before the arch). The dotted line plotted vertically on the x-axis is at 0, denoting the marginal stability of $N_{\omega,\mathrm{sp}} = 0$. The dotted line plotted horizontally on the y-axis is at 0.24 cm/year, the empirically found growth threshold that optimally discriminates between stable and unstable aneurysms predicted by the proposed aneurysm physiomarker.

### S1.11. **Growth in SOV and MAA vs local aneurysm physiomarker.**

S1.12. **ROC analysis.** The optimal cut-off was found as the point where the ROC curve crosses a straight line with slope $g$, such that

$$g = \frac{\text{costfn}(P|N) - \text{costfn}(N|N)}{\text{costfn}(N|P) - \text{costfn}(P|P)} \frac{y}{x}. \tag{S11}$$

Here, x is the number of observations in the positive classification, and y is the total number of observations in the negative classification. The cost function gives components of the matrix B

$$B = \begin{bmatrix} \text{costfn}(P|P) & \text{costfn}(N|P) \\ \text{costfn}(P|N) & \text{costfn}(N|N) \end{bmatrix} = \begin{bmatrix} 0 & 1 \\ 1 & 0 \end{bmatrix}, \tag{S12}$$

where costfn(N|P) is the cost of misclassifying a positive classification as a negative classification and vice versa. The optimal operating point is found at the intersection of the straight line with slope m from the upper left corner of the ROC plot (sensitivity=specificity=1) and the ROC curve.

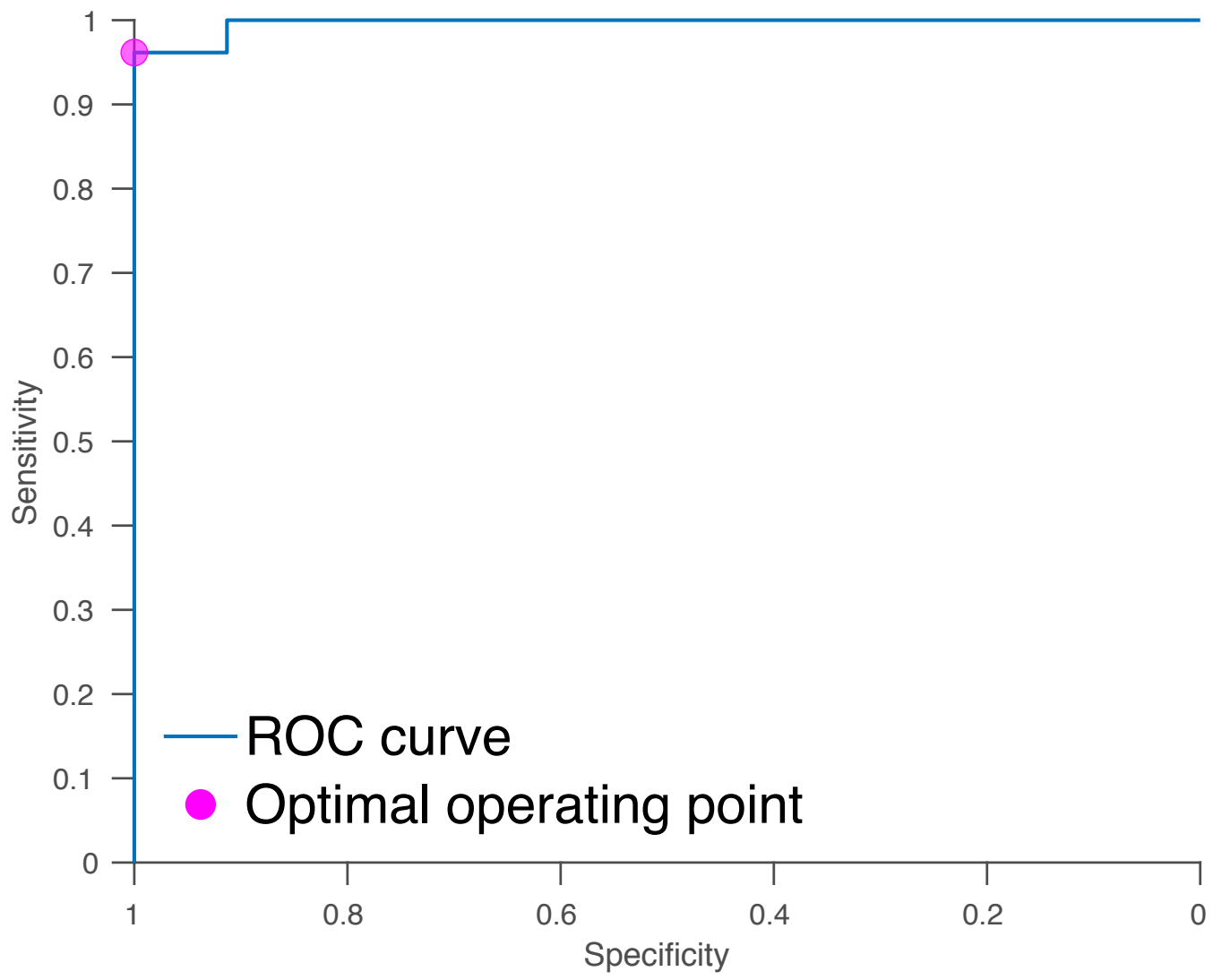

FIGURE S7. The receiver operating characteristic (ROC) analysis curve, plotted as sensitivity vs. specificity across varying cutoffs of $N_{\omega,\text{sp}}$. The optimal operating point[18] (sensitivity = 0.96, specificity = 1) occurs at the $N_{\omega,\text{sp}} = 0.07 \approx 0$ for patients with follow-up data, suggesting that the analytically derived threshold $N_{\omega,\text{threshold}}$ accurately describes the onset of the underlying instability.

S1.13. **Establishing the marginal stability curve.** From the characteristic equations 22 and 23, we can determine the marginal stability curve where $\tilde{\mu} = 0$ via the method proposed by Kumar et al[19]. The measurable values of $k''$, $\tilde{\omega}$, and $N_m$ are fixed for a specific flow scenario, yielding an eigenvalue problem for the critical $N_{\omega,\text{threshold}}$ associated with $\tilde{\mu} = 0$.

We first write our solution set of Fourier coefficients $\tilde{A}_{k,n} = \tilde{A}^r_{k,n} + i\tilde{A}^i_{k,n}$, $u''_{k,n} = u''^r_{k,n} + iu''^i_{k,n}$ in terms of real and imaginary components. Then the dimensionless characteristic equations 22 and 23 can likewise be separated into purely real and imaginary parts

**Mass equation**

$$\begin{aligned}
&\tilde{A}^r_{k,n}\mu + \tilde{A}^i_{k,n}\left(-\tilde{\omega}(\alpha+n) - \frac{k''N_m}{2}\right) + \frac{\tilde{A}^r_{k,n-1}k''N_\omega(2\text{ph}[\beta_b]^i + \tilde{\omega})}{8\text{ph}[\beta_b]^{r2} + 2(2\text{ph}[\beta_b]^i + \tilde{\omega})^2} \\
&+\frac{\tilde{A}^r_{k,n+1}k''N_\omega(2\text{ph}[\beta_b]^i - \tilde{\omega})}{8\text{ph}[\beta_b]^{r2} + 2(\tilde{\omega} - 2\text{ph}[\beta_b]^i)^2} - \frac{\tilde{A}^i_{k,n-1}\text{ph}[\beta_b]^r k''N_\omega}{4\text{ph}[\beta_b]^{r2} + (2\text{ph}[\beta_b]^i + \tilde{\omega})^2} \\
&-\frac{\tilde{A}^i_{k,n+1}\text{ph}[\beta_b]^r k''N_\omega}{4\text{ph}[\beta_b]^{r2} + (\tilde{\omega} - 2\text{ph}[\beta_b]^i)^2} - k''u''^i_{k,n} \\
&+i\left(\tilde{A}^r_{k,n}\left(\tilde{\omega}(\alpha+n) + \frac{k''N_m}{2}\right) + \frac{\tilde{A}^r_{k,n-1}\text{ph}[\beta_b]^r k''N_\omega}{4\text{ph}[\beta_b]^{r2} + (2\text{ph}[\beta_b]^i + \tilde{\omega})^2} + \frac{\tilde{A}^r_{k,n+1}\text{ph}[\beta_b]^r k''N_\omega}{4\text{ph}[\beta_b]^{r2} + (\tilde{\omega} - 2\text{ph}[\beta_b]^i)^2}\right. \\
&\left.+\tilde{A}^i_{k,n}\mu + \frac{\tilde{A}^i_{k,n-1}k''N_\omega(2\text{ph}[\beta_b]^i + \tilde{\omega})}{8\text{ph}[\beta_b]^{r2} + 2(2\text{ph}[\beta_b]^i + \tilde{\omega})^2} + \frac{\tilde{A}^i_{k,n+1}k''N_\omega(2\text{ph}[\beta_b]^i - \tilde{\omega})}{8\text{ph}[\beta_b]^{r2} + 2(\tilde{\omega} - 2\text{ph}[\beta_b]^i)^2} + k''u''^r_{k,n}\right) = 0,
\end{aligned} \tag{S13}$$

**Momentum equation**

$$\begin{aligned}
&-\frac{\tilde{A}^r_{k,n}\beta_m N_m}{|\beta_b|} + u''^i_{k,n}\left(-\frac{2\beta^i_p}{|\beta_b|} - \tilde{\omega}(\alpha+n) - \frac{k''N_m}{2}\right) + u''^r_{k,n}\left(\frac{2\beta^r_p}{|\beta_b|} + \mu\right) \\
&-\frac{\tilde{A}^r_{k,n-1}N_\omega\left(2\text{ph}[\beta_b]^{r2} + \text{ph}[\beta_b]^i(2\text{ph}[\beta_b]^i + \tilde{\omega})\right)}{4\text{ph}[\beta_b]^{r2} + (2\text{ph}[\beta_b]^i + \tilde{\omega})^2} + \frac{\tilde{A}^r_{k,n+1}N_\omega\left(\text{ph}[\beta_b]^i\tilde{\omega} - 2\left(c^2 + d^2\right)\right)}{4\text{ph}[\beta_b]^{r2} + (\tilde{\omega} - 2\text{ph}[\beta_b]^i)^2} - \tilde{A}^i_{k,n}k \\
&-\frac{\tilde{A}^i_{k,n-1}\text{ph}[\beta_b]^r N_\omega\tilde{\omega}}{4\text{ph}[\beta_b]^{r2} + (2\text{ph}[\beta_b]^i + \tilde{\omega})^2} + \frac{\tilde{A}^i_{k,n+1}\text{ph}[\beta_b]^r N_\omega\tilde{\omega}}{4\text{ph}[\beta_b]^{r2} + (\tilde{\omega} - 2\text{ph}[\beta_b]^i)^2} + \frac{k''N_\omega u''^r_{k,n-1}(2\text{ph}[\beta_b]^i + \tilde{\omega})}{8\text{ph}[\beta_b]^{r2} + 2(2\text{ph}[\beta_b]^i + \tilde{\omega})^2} \\
&+\frac{k''N_\omega u''^r_{k,n+1}(2\text{ph}[\beta_b]^i - \tilde{\omega})}{8\text{ph}[\beta_b]^{r2} + 2(\tilde{\omega} - 2\text{ph}[\beta_b]^i)^2} - \frac{\text{ph}[\beta_b]^r k''N_\omega u''^i_{k,n-1}}{4\text{ph}[\beta_b]^{r2} + (2\text{ph}[\beta_b]^i + \tilde{\omega})^2} - \frac{\text{ph}[\beta_b]^r k''N_\omega u''^i_{k,n+1}}{4\text{ph}[\beta_b]^{r2} + (\tilde{\omega} - 2\text{ph}[\beta_b]^i)^2} \\
&+i\left(\tilde{A}^r_{k,n}k + u''^r_{k,n}\left(\frac{2\beta^i_p}{|\beta_b|} + \tilde{\omega}(\alpha+n) + \frac{k''N_m}{2}\right) - \frac{\tilde{A}^i_{k,n}\beta_m N_m}{|\beta_b|} + u''^i_{k,n}\left(\frac{2\beta^r_p}{|\beta_b|} + \mu\right)\right. \\
&+\frac{\tilde{A}^r_{k,n-1}\text{ph}[\beta_b]^r N_\omega\tilde{\omega}}{4\text{ph}[\beta_b]^{r2} + (2\text{ph}[\beta_b]^i + \tilde{\omega})^2} - \frac{\tilde{A}^r_{k,n+1}\text{ph}[\beta_b]^r N_\omega\tilde{\omega}}{4\text{ph}[\beta_b]^{r2} + (\tilde{\omega} - 2\text{ph}[\beta_b]^i)^2} \\
&-\frac{\tilde{A}^i_{k,n-1}N_\omega\left(2\text{ph}[\beta_b]^{r2} + \text{ph}[\beta_b]^i(2\text{ph}[\beta_b]^i + \tilde{\omega})\right)}{4\text{ph}[\beta_b]^{r2} + (2\text{ph}[\beta_b]^i + \tilde{\omega})^2} + \frac{\tilde{A}^i_{k,n+1}N_\omega\left(\text{ph}[\beta_b]^i\tilde{\omega} - 2\left(c^2 + d^2\right)\right)}{4\text{ph}[\beta_b]^{r2} + (\tilde{\omega} - 2\text{ph}[\beta_b]^i)^2} \\
&+\frac{\text{ph}[\beta_b]^r k''N_\omega u''^r_{k,n-1}}{4\text{ph}[\beta_b]^{r2} + (2\text{ph}[\beta_b]^i + \tilde{\omega})^2} + \frac{\text{ph}[\beta_b]^r k''N_\omega u''^r_{k,n+1}}{4\text{ph}[\beta_b]^{r2} + (\tilde{\omega} - 2\text{ph}[\beta_b]^i)^2} \\
&\left.+\frac{k''N_\omega u''^i_{k,n-1}(2\text{ph}[\beta_b]^i + \tilde{\omega})}{8\text{ph}[\beta_b]^{r2} + 2(2\text{ph}[\beta_b]^i + \tilde{\omega})^2} + \frac{k''N_\omega u''^i_{k,n+1}(2\text{ph}[\beta_b]^i - \tilde{\omega})}{8\text{ph}[\beta_b]^{r2} + 2(\tilde{\omega} - 2\text{ph}[\beta_b]^i)^2}\right) = 0,
\end{aligned} \tag{S14}$$

where $\text{ph}[\beta_b] = \text{ph}[\beta_b]^r + i\text{ph}[\beta_b]^i$ and $\beta_b = \beta_b^r + i\beta_b^i$. The real and imaginary parts of eqn. S13 as well as S14 must each be identically zero to fully satisfy both characteristic equations. This provides four equations for four unknowns, $\underline{Z} = [\tilde{A}^r_{k,n}, \tilde{A}^i_{k,n}, u''^r_{k,n}, u''^i_{k,n}]'$, which can be written as a linear equation $\boldsymbol{J}\underline{Z} = 0$. The matrix $\boldsymbol{J}$ thus comprises the linear coefficients of $\underline{Z}$ in the real & imaginary parts of the mass (eqn. S13) and momentum (eqn. S14) equations.

The real part of the temporal growth rate is set to $\tilde{\mu} = 0$. The imaginary part takes on the value $\alpha = 1/2$ for the subharmonic resonance in which the response frequency is half the driving frequency, and $\alpha = 0$ for the harmonic case where the response frequency is the same as the driving frequency $\tilde{\omega}$. Since integer multiples of $\tilde{\omega}$ can be absorbed into the periodic function $\boldsymbol{P}(t)$, $\alpha$ is defined modulo $\tilde{\omega}$. The Fourier terms with $0 < \alpha < 1/2$ is equivalent to the complex conjugate terms associated with $1/2 < \alpha < 1$, so consideration can be restricted to $0 \leq \alpha \leq 1/2$. However, we note that $0 < \alpha < 1/2$ are only associated with stable flows $\tilde{\mu} \leq 0$ in the range of physiologically viable $N_\omega$ [19]. Kumar et al. showed in their analogous analysis of an interface between two fluids that only the harmonic and subharmonic cases are relevant to the linear stability problem. The complex Floquet multipliers associated with $0 \leq \alpha \leq 1/2$ correspond to damped, stable solutions. We have verified that this holds true for selected $\alpha$ in the range $0 \leq \alpha \leq 1/2$ for physiologically viable $N_\omega$, though a rigorous theoretical proof remains an open problem.

The critical $N_{\omega,\text{crit}}$ corresponding to the marginal stability curve $\tilde{\mu} = 0$ as well as the $k''$, $\tilde{\omega}$, and $N_m$ selected for a specific flow scenario can be found by solving an eigenvalue problem. Specifically, we decompose $\boldsymbol{J}$ into $\boldsymbol{C}$, the linear coefficients of $\underline{Z}$ in $\boldsymbol{J}$ that do not contain $N_\omega$, and $\boldsymbol{D}$, the entries of $\boldsymbol{J}$ that are proportional to $N_\omega$. The linear matrix equation becomes

$$\boldsymbol{C}\underline{Z} + N_\omega \boldsymbol{D}\underline{Z} = 0. \tag{S15}$$

This can be written as an eigenvalue problem

$$-\text{inv}(\boldsymbol{C})\boldsymbol{D}\underline{Z} = \frac{1}{N_{\omega,\text{crit}}}\underline{Z}, \tag{S16}$$

where the eigenvalues of the matrix $-\text{inv}(\boldsymbol{C})\boldsymbol{D}$ are reciprocals of the critical $N_{\omega,\text{crit}}$ on the marginal stability curve. The preset values of $\tilde{\omega}$ and $N_m$ are measured from patient MRI, and $k''$ is swept through to obtain the "tongues" observed in Fig. 2. The minimum critical $N_\omega$ associated with the first subharmonic tongue to appear as $k''$ increases is chosen as the threshold $N_{\omega,\text{threshold}}$. It is the global minimum of $N_{\omega,\text{crit}}$ on all marginal stability tongues, such that the flutter instability is triggered first for increasing $N_\omega$ at the value $N_{\omega,\text{threshold}}$.